\documentclass[paper]{JHEP3}
\usepackage{cite}
\usepackage{amssymb,amsmath}
\usepackage{graphicx}
\usepackage{xspace}
\usepackage{booktabs}

\graphicspath{{figures/}}

\def\Q2{\left(Q^{2}\right)}
\def\e{\epsilon}

\def\ep{\epsilon}
\def\d{{\rm d}}

\def\Li{\hbox{Li}}

\def\l({\left(}
\def\r){\right)}

\def\e{\epsilon}

\def\d{\hbox{d}}

\def\l{(}
\def\r{)}

\newcommand{\sigmahat}{\ensuremath{\hat{\sigma}}}

\newcommand{\diff}{\ensuremath{\text{d}}}

\newcommand{\dsigma}{\ensuremath{\diff{\sigmahat}}}

\newcommand{\cross}[1]{ \widehat{ #1 } }
\newcommand{ \q }{ q }
\newcommand{ \qb }{ \overline{q} }
\newcommand{ \gluon }{ g }
\newcommand{ \qp }{ q' }
\newcommand{ \qpb }{ \overline{q}' }
\newcommand{ \qbp }{ \qpb }
\newcommand{ \limfreemark }{*}

\newcommand{ \imom }[1]{ p_{#1} }
\newcommand{ \pa }{ \imom{1} }
\newcommand{ \pb }{ \imom{2} }
\newcommand{ \fmom }[1]{ k_{#1} }
\newcommand{ \fmomenta }{ \fmom{1}, \ldots, \fmom{m+2} }
\newcommand{ \psmomenta }{ \fmomenta ; \pa, \pb }
\newcommand{ \psmomentaini }{ \pa, \pb; \fmomenta }
\newcommand{ \dmom }[1]{ \left[ \d \fmom{#1} \right] }

\newcommand{ \reduced }[1]{ \tilde{ #1 } }
\newcommand{ \fmomred }[1]{ \reduced{ \fmom{} }_{#1} }
\newcommand{ \fmomentared }{ \fmomred{1}, \dotsc, \fmomred{i}, \fmomred{l}, \dotsc, \fmomred{m+2} }
\newcommand{ \xa }{ x_1 }
\newcommand{ \xb }{ x_2 }
\newcommand{ \xahat }{ \hat{x}_1 }
\newcommand{ \xbhat }{ \hat{x}_2 }

\newcommand{ \secRef }[1]{Section~\ref{#1}\xspace}

\title{Antenna subtraction at NNLO with hadronic initial states:
double real initial-initial configurations}
\author{A.\ Gehrmann-De Ridder$^a$, T.\ Gehrmann$^{b}$, M.\ Ritzmann$^c$\\
$^a$ Institute for Theoretical Physics, ETH, CH-8093 Z\"urich, Switzerland\\
$^b$ Institut f\"ur Theoretische Physik, Universit\"at Z\"urich,
Wintherturerstrasse 190,\\CH-8057 Z\"urich, Switzerland\\
$^c$  Institut de Physique Th\'{e}orique, CEA-Saclay, F-91191 Gif-sur-Yvette cedex, France}

\abstract{
The antenna subtraction method handles 
real radiation contributions in higher order corrections to jet observables. 
The method is based on antenna functions, which encapsulate all 
unresolved radiation between a pair of hard radiator partons. To apply this method to 
compute hadron collider observables, 
initial-initial antenna functions with both radiators in the initial
state are required in unintegrated and integrated forms. 
In view of extending the antenna subtraction method to next-to-next-to-leading order 
(NNLO) calculations at hadron colliders, we derive the 
full set of initial-initial double real radiation antenna functions in integrated form. }

\preprint{{ZU-TH 15/12}\\{Saclay-IPhT-T12/061}}

\begin{document}
%\maketitle
\allowdisplaybreaks

\section{Introduction}

Jet production at high energy colliders allows to reconstruct the parton-level dynamics 
of the hard interaction processes. Jet cross sections are therefore ideally suited for 
precision studies of QCD~\cite{dissertori} to determine the strong coupling constant or
parton distribution functions. They can be measured to a very high experimental accuracy, 
which has to be met by an equally high theoretical precision in order to perform meaningful 
comparisons. This precision demands the inclusion of next-to-next-to-leading order 
(NNLO) QCD corrections. 

NNLO calculations of observables with $n$ jets in the final state require
several ingredients: the two-loop corrected $n$-parton matrix elements, the
one-loop corrected $(n+1)$-parton matrix elements, and the tree-level
$(n+2)$-parton matrix elements. For many massless jet observables of
phenomenological interest, these matrix elements are available for some time 
already. 

The $(n+1)$-parton and $(n+2)$-parton matrix elements 
contribute to $n$ jet observables at NNLO if the extra partons are unresolved or
are clustered to form an
$n$-jet final state. Consequently, these extra partons are unconstrained in the 
soft and collinear regions, and  yield infrared divergences. 
In these cases, the infrared
singular parts of the matrix elements need to be extracted and integrated
over the phase space appropriate to the unresolved configuration
to make the infrared pole structure explicit. The single soft and collinear limits of 
one-loop matrix elements~\cite{onelstr,oneloopsoft,onelstr1,onelstr2,twolstr} and the  
double unresolved limits of tree-level matrix 
elements~\cite{audenigel,campbell,cg,campbellandother} are process-independent, 
and result in a factorisation into an unresolved factor times a matrix element 
of lower multiplicity.

To determine the contribution to NNLO jet observables from these configurations, one has to
find  subtraction terms which coincide with the full matrix element
in the unresolved limits
and are still sufficiently simple to be integrated analytically in order
to cancel  their  infrared pole structure with the virtual contributions. 
Often starting from systematic
methods for subtraction at NLO~\cite{kunszt,cs,ant,nlosub},
several NNLO subtraction methods have been
proposed in the
literature~\cite{Kosower:2002su,nnlosub2,nnlosub3,nnlosub4,nnlosub5,nnlosub6,nnlosub7, nnlosub8}, and are
worked out
to a varying level of sophistication.

For observables with partons only in the final state,
an NNLO subtraction formalism, antenna subtraction,
 has been derived in~\cite{ourant}. The antenna subtraction formalism
constructs the subtraction terms from antenna functions. Each
antenna function encapsulates all singular limits due to the
emission of one or two unresolved partons between two colour-connected
hard radiator partons. This construction exploits the  universal factorisation
of matrix elements and phase space in all unresolved limits.
The antenna functions are derived
systematically from physical matrix elements~\cite{our2j}. This formalism
has been applied in the derivation of NNLO corrections to three-jet
production in electron-positron annihilation~\cite{our3j,weinzierljet}
and related event shapes~\cite{ourevent,weinzierlevent}, which
were used subsequently in precision determinations of the strong coupling
constant~\cite{ouras,bechernew,hoang,davisonwebber,bethkeas,power}.
The formalism
can be extended to include massive fermions~\cite{ritzmann}.
Antenna subtraction has also been used as the starting point for a 
parton shower algorithm \cite{antshower}, using
 the fact that antennae capture the unresolved limits of QCD 
amplitudes to achieve a leading-log resummation.

For processes
with initial-state partons, antenna subtraction has been fully worked out
only to NLO so far~\cite{hadant}. In this case, one encounters two new types
of antenna functions, initial-final antenna functions with one radiator parton
in the initial state, and initial-initial antenna functions with both
radiator partons in the initial state. The framework for the construction of 
NNLO antenna subtraction 
terms involving one or two partons in the initial state has been set up in~\cite{joao,joaonew}
in the context of a proof-of-principle implementation of the contribution of the 
purely gluonic contributions to  di-jet production at hadron colliders. The initial-final  and 
initial-initial antenna functions appearing in the NNLO subtraction terms are 
obtained from crossing the final-final antennae. Their integration has to be performed 
over the appropriate phase space. In the case of the initial-final antennae, this 
has been accomplished in~\cite{gionata}. The initial-initial one-loop antenna functions 
were integrated in~\cite{monni2}.  
 Partial results have been 
obtained for the integrated initial-initial double real radiation tree-level antenna functions  
in~\cite{mathias},
where all antennae involving a secondary quark-antiquark pair were integrated. 
 It is the aim of the present paper to complete the
NNLO antenna subtraction scheme for hadron collider processes by integrating the remaining 
initial-initial double real radiation antenna functions. 

Other approaches to perform NNLO calculations  of exclusive observables with initial
state partons are  the use of sector decomposition (partly in combination with subtraction)
 and a subtraction method based on the
transverse momentum structure of the final state.  The sector decomposition
algorithm~\cite{secdec}  analytically decomposes both phase space and loop integrals into
their  Laurent expansion in dimensional regularisation, and  performs a subsequent
numerical  computation of the coefficients of this expansion. Using this formalism,  NNLO
results were  obtained for Higgs  production~\cite{babishiggs}     and vector boson
production~\cite{babisdy} at hadron colliders. Both reactions  were equally computed
independently~\cite{grazzinihiggs} using an NNLO subtraction  formalism exploiting the specific transverse
momentum  structure of these observables~\cite{nnlosub6}, which was also
applied
to compute NNLO corrections to associated $WH$ production~\cite{grazziniwh} and 
photon pair production~\cite{grazzinigg}.
A very promising approach could be
the combination of 
subtraction with sector decomposition~\cite{nnlosub7}, which was recently applied in the 
calculation of NNLO corrections to the quark-induced processes in top quark pair 
production~\cite{czakonmitov}.

This paper is structured as follows: in Section~\ref{sec:ant}, we
briefly summarise the application of the antenna subtraction formalism to hadronic 
collisions, which was developed in~\cite{hadant,mathias,joao,joaonew}. The 
antenna functions required for double real radiation and their associated phase
space mappings are described in Section~\ref{sec:phase} while the phase space integration 
is described in Section~\ref{sec:int}.
The resulting integrated initial-initial antenna functions 
being too lengthy to be presented here, only their leading pole parts are
given in Section~\ref{sec:results}. The full results are attached
separately to this paper as a FORM~\cite{FORM} file. 
Section~\ref{sec:conc} contains our conclusions and an 
outlook. The newly derived phase space master integrals required for our 
calculation are collected in the Appendix.

 \section{Initial-initial antenna subtraction at NNLO}
 \label{sec:ant}
 The partonic double real contribution to an NNLO $m$-jet cross section reads
 \begin{eqnarray}
\lefteqn{{\rm d}\hat\sigma^{RR}_{NNLO}
= {\cal N}^{RR}_{NNLO} \,\sum_{\text{perms}}{\rm d}\Phi_{m+2}( \psmomenta )
\frac{1}{S_{{m+2}}} }\nonumber \\ &\times&
|{\cal M}_{m+2}( \psmomentaini )|^{2}\;
J_{m}^{(m+2)}( \psmomenta )\;.
\label{eq:nnloreal}
\end{eqnarray}
In this equation,
$|{\cal M}_{m+2}( \psmomentaini )|^{2}$ stands for the colour-ordered 
$2 \rightarrow m+2$ amplitude norm squared (or, for subleading 
colour contributions, for the appropriate sum of interference terms between 
colour-ordered amplitudes).
The symmetry factor $S_{m+2}$ accounts for identical partons in the
final state and 
$\sum_{\text{perms}}$ denotes the sum over all configurations 
with $m+2$ partons. 
The next-to-next-to leading order normalisation factor 
${\cal N}^{RR}_{NNLO}$ includes all QCD-independent factors 
as well as the dependence on the renormalised QCD coupling constant
$\alpha_s$. It is related to the normalisation factor present at
leading order, ${\cal N}_{LO}$, which depends 
on the specific process and parton channel under consideration by, 
\begin{equation}
{\cal N}^{RR}_{NNLO}={\cal N}_{LO}  \left(\frac{\alpha_s N}{2\pi}\right)^{2}
\frac{\bar{C}(\epsilon)^2}{C(\epsilon)^2},\\
\end{equation}
where, 
\begin{eqnarray}
\label{eq:Cdef}
C(\epsilon)=(4\pi)^{\epsilon}\frac{e^{-\epsilon\gamma}}{8\pi^2},\\
\label{eq:Cbar}
\bar{C}(\e)=(4\pi)^{\e}e^{-\e\gamma}.
\end{eqnarray}

The initial state momenta are labelled as usual as $\pa$ and $\pb$ whereas the $m+2$ 
 momenta in the final state are labelled $\fmomenta$. 
The $2\to m+2$ particle phase space is denoted as
\begin{eqnarray}
{\rm d}\Phi_{m+2}( \psmomenta )= \dmom{1}
\hdots 
\dmom{m+2} (2\pi)^d
\delta^d( \pa + \pb - \fmom{1} -\hdots- \fmom{m+2})
\end{eqnarray}
where we have introduced the abbreviation $ \dmom{l} = \diff^{d} \fmom{l} \delta^+ ( \fmom{l}^2)/ (2 \pi)^{d-1}$.
The jet function $J_{m}^{(m+2)}( \psmomenta )$ 
ensures that out of ($m+2$) final state partons, 
an observable with $m$ jets is built. 
The incoming parton momenta $\pa, \pb$ serve as reference directions to define transverse momenta and rapidities of the jets.
 
 The real radiation contribution (\ref{eq:nnloreal}) contains infrared divergencies, which arise 
 when one or two of the final state partons become unresolved. The numerical phase 
 space integration in~(\ref{eq:nnloreal}) can therefore not be carried out without prior 
 introduction of a regulator or a subtraction. A subtraction term ${\rm d}\hat\sigma^{S}_{NNLO}$ 
is defined on the same phase space as ${\rm d}\hat\sigma^{RR}_{NNLO}$, and it approaches its 
integrand in all unresolved limits. Consequently 
${\rm d}\hat\sigma^{RR}_{NNLO}-{\rm d}\hat\sigma^{S}_{NNLO}$ is finite and the integration 
can be performed numerically. ${\rm d}\hat\sigma^{S}_{NNLO}$ is integrated over those parts 
of the phase space that contain infrared singularities and then
combined with the virtual NNLO and the mass factorisation counterterm 
contributions to achieve the cancellation of infrared divergencies. 

The double real radiation contribution $\dsigma_{NNLO}^{RR}$
can become singular if either one or two final state partons 
are unresolved (soft or collinear).
Consequently, when constructing the corresponding subtraction term 
$\dsigma_{NNLO}^{S}$ in the antenna subtraction method, 
we must distinguish the following 
configurations according to the colour connection of the unresolved partons:
\begin{itemize}
\item[(a)] One unresolved parton but the experimental observable selects only
$m$ jets.
\item[(b)] Two colour-connected unresolved partons (colour-connected).
\item[(c)] Two unresolved partons that are not colour connected but share 
a common radiator (almost colour-unconnected).
\item[(d)] Two unresolved partons that are well separated from each other 
in the colour 
chain (colour-unconnected).
\item[(e)] Large angle soft gluon radiation. 
\end{itemize}

All cases except (b) and (e) 
can be accounted for by combinations of three-parton antenna functions, 
which were derived and integrated already in the context of antenna subtraction at NLO. 
Case (e) can always be described~\cite{joaonew}
by soft antenna functions in final-final kinematics, that were derived in~\cite{our3j,weinzierljet}.  
In case (b), tree-level four-parton antennae are needed.
The corresponding subtraction term for both radiators in the initial state reads:
\begin{eqnarray}
\lefteqn{{\rm d}\hat\sigma_{NNLO}^{S,b,(ii)}
=  {\cal N}^{RR}_{NNLO}\,\sum_{m+2}{\rm d}\Phi_{m+2}( \psmomenta )
\frac{1}{S_{m+2}} }
\nonumber \\
&\times& \,\sum_{il} \sum_{jk}\;\left( X^0_{il,jk}
- X^0_{l,jk} X^0_{iL,K} - X^0_{i,kj} X^0_{Il,J} \right)
|{\cal M}_{m}( x_i p_i, x_l p_l ; \fmomentared )|^2
\nonumber \\
&\times&
J_{m}^{(m)}( \fmomentared ; x_i p_i, x_l p_l )\;.
\label{eq:sub2bii}
\end{eqnarray}
The sum runs over all colour-adjacent pairs $j,k$ and implies that 
the hard momenta $i,l$ are chosen accordingly. 
Tree-level antenna functions are generically denoted $X^0$, we will 
give the precise definition in \secRef{sec:ant}. 
$X^0_{il,jk}$ contains all the unresolved limits 
associated with $j$ and/or $k$ becoming becoming unresolved between the 
initial-state emitters $i$ and $l$.
Its single unresolved limits are subtracted by products of three-particle 
antenna functions, ensuring that $\diff \sigmahat_{NNLO}^{S,b,(ii)}$ is only
active in the double unresolved region.
The reduced matrix element ${\cal M}$ is evaluated with a set of $m$ 
on-shell momenta which are obtained from the original ones by rescaling 
the initial-state momenta $p_i$ and $p_l$, omitting the final-state momenta 
$\fmom{j}$ and $\fmom{k}$ and applying a Lorentz boost to the remaining 
final-state momenta~\cite{hadant}.

To analytically integrate the subtraction term (\ref{eq:sub2bii}), we employ dimensional 
regularisation in $d=4-2\e$ dimensions. 
We use the factorisation of the 
phase space
\begin{eqnarray}
\label{eq:psii4}
{\rm d}\Phi_{m+2}( \psmomenta )&=&
{\rm d}\Phi_{m}( \fmomentared ; \xa \pa, \xb \pb )
\nonumber\\
&&\times\delta( \xa - \xahat )\,\delta( \xb - \xbhat ) \dmom{j} \dmom{k} \d \xa \d \xb\,,
\end{eqnarray}
where $\xahat$ and $\xbhat$ are defined throughout the phase space. In the case of 
collinear initial state radiation, they are identified with the collinear momentum 
fractions of the partons entering the hard scattering process. 

The kinematics of the 
reduced matrix element appearing in (\ref{eq:sub2bii}) depends on the mapped momenta,
and thus consequently  
on $q^2,\xa,\xb$. The integration of the antenna functions must therefore retain the dependence 
on $\xa,\xb$. The integrated initial-initial antenna functions are defined as:
 \begin{eqnarray}
\label{eq:x4intii}
&&{\cal X}^0_{il,jk}(x_i,x_l, \ep)=\frac{1}{[C(\epsilon)]^2}\int \dmom{j} \dmom{k} \;x_i\;x_l\; \delta(x_i-\hat{x}_i)\,\delta(x_l-\hat{x}_l)\,X^0_{il,jk},\,
\end{eqnarray}
where $C(\epsilon)$ is given in eq.~(\ref{eq:Cdef}).

\section{Antenna functions for double real radiation}
\label{sec:phase}

\subsection{Definition of the antenna functions}
Antenna functions are characterised by their parton content and their radiators, 
i.e.~the pair of hard partons which they collapse to in the unresolved limits.
Accordingly, we group them into quark-antiquark, quark-gluon and gluon-gluon antennae. 
They are derived from physical matrix 
elements associated to the decay of a colourless particle into partons~\cite{our2j}. 
The  tree-level antenna functions are obtained by normalising the three- and four-parton tree-level 
colour sub-amplitudes squared to that of the basic two-parton process:
The final-final three- and four-particle antennae are respectively defined by:
\begin{eqnarray}
X_{ijk}^0 = S_{ijk,IK}\, \frac{|{\cal M}^0_{ijk}|^2}{|{\cal M}^0_{IK}|^2}\;,\nonumber\\
X_{ijkl}^0 = S_{ijkl,IL}\, \frac{|{\cal M}^0_{ijkl}|^2}{|{\cal M}^0_{IL}|^2}\;.
\end{eqnarray}
where $S$ denotes the symmetry factor associated with the antenna, which accounts
both for potential identical particle symmetries and for the presence 
of more than one antenna in the basic two-parton process.
It is chosen such that the antenna function reproduces the unresolved 
limits of a matrix element with identified particles.
The initial-initial tree-level three- and four-parton antennae denoted 
by $X_{ik,l}^0$ and $X_{il,jk}^0$ are obtained by crossing two partons 
to the initial state, starting from the final-final antennae. The distinct crossings of the three-parton 
antenna functions are listed in Table~\ref{tab:nlo}.
Crossings which are free of unresolved limits are marked with a \limfreemark \limfreemark.
\begin{table}[t]
\begin{center}
\begin{tabular*}{0.9 \textwidth}{l @{$\qquad$} l}
\toprule
\multicolumn{2}{l}{quark-antiquark antennae}\\
\midrule
$A^0_3$	&
$A^0_3 \big( \cross{ \q },	\cross{ \gluon }, 	\qb		 	\big)$,
$A^0_3 \big( \cross{ \q },	\gluon,			\cross{\qb}	\big)$
\\[3pt]
\midrule
\multicolumn{2}{l}{quark-gluon antennae}\\
\midrule
$D^0_3$ &
$D^0_3 \big( \cross{ \q },	\cross{ \gluon }, 	\gluon	 	\big)$,
$D^0_3 \big( \q,		\cross{ \gluon }, 	\cross{\gluon}	 \big)$
\\
$E^0_3$ &
$E^0_3 \big( \cross{ \q },	\cross{ \qp }, 	\qbp	 	\big)$,
$E^0_3 \big( \q,		\cross{ \qp }, 	\cross{ \qbp }	 \big)^{\limfreemark \limfreemark}$
\\[3pt]
\midrule
\multicolumn{2}{l}{gluon-gluon antennae}\\
\midrule
$F^0_3$ &
$F^0_3 \big( \cross{ \gluon },	\gluon, 	\cross{ \gluon }	 	\big)$\\
$G^0_3$ &
$G^0_3 \big( \cross{ \gluon },	\q, 		\cross{ \qb }	 	\big)$,
$G^0_3 \big( \gluon,			\cross{\q}, 	\cross{ \qb }	 	\big)^{\limfreemark \limfreemark}$\\
\bottomrule
\end{tabular*}
\end{center}
\caption{
The tree-level three-particle initial-initial antennae and their distinct crossings.
$D^0_3$ is symmetric under the interchange of the two gluons, $F^0_3$ 
under cyclic interchange of its arguments.
\label{tab:nlo}}
\end{table}

The four-particle tree-level antenna functions are not determined 
by the species of the particles alone but also by the colour-connection.
We distinguish leading-colour antennae, denoted by letters without tilde, 
where the particles are colour-connected in the order they are listed 
and subleading colour antennae, denoted by letters with tilde, where the gluons 
are photon-like. This notation has been established in~\cite{ourant,gionata}.
 The unresolved limits of the initial-initial antennae can be obtained from 
those of the final-final antennae by crossing. 
The crossing of the triple-collinear splitting functions 
is explained in \cite{dfg}.

\begin{table}[t]
\begin{center}
\begin{tabular*}{0.9 \textwidth}{l @{$\qquad$} l}
\toprule
\multicolumn{2}{l}{quark-antiquark antennae}\\
\midrule
$A^0_4$	&
$A^0_4 \big( \cross{ \q },	\gluon,			\gluon,			\cross{\qb}	\big)$,
$A^0_4 \big( \cross{ \q },	\cross{ \gluon }, 	\gluon,			\qb		 	\big)$,
$A^0_4 \big( \cross{ \q },	\gluon,			\cross{\gluon},		\qb			\big)$,
$A^0_4 \big( \q,			\cross{\gluon},		\cross{\gluon},		\qb			\big)$
\\[3pt]
$\widetilde{A}^0_4$	&
$\widetilde{A}^0_4 \big( \cross{ \q },	\gluon,			\gluon,			\cross{\qb}	\big)$,
$\widetilde{A}^0_4 \big( \cross{ \q },	\cross{ \gluon }, 	\gluon,			\qb		 	\big)$,
$\widetilde{A}^0_4 \big( \q,		\cross{\gluon},		\cross{\gluon},		\qb			\big)$
\\[3pt]
$B^0_4$	&
$B^0_4 \big( \cross{ \q },	\qp, 			\qpb,			\cross{ \qb } 	\big)$,
$B^0_4 \big( \cross{ \q },	\cross{ \qp }, 	\qpb,			\qb 			\big)$,
$B^0_4 \big( \q,		\cross{ \qp }, 	\cross{ \qpb },	\qb
\big)^{\limfreemark \limfreemark}$
\\[3pt]
$C^0_4$	&
$C^0_4 \big( \cross{ \q },	\qb,			\q,			\cross{\qb}	\big)$,
$C^0_4 \big( \cross{ \q },	\qb, 			\cross{ \q},		\qb	 		\big)$,
$C^0_4 \big( \q,		\cross{ \qb },	\cross{ \q },	\qb
\big)^{\limfreemark \limfreemark}$,
$C^0_4 \big( \q,		\cross{ \qb },			 \q, \cross{ \qb }	\big)^{\limfreemark \limfreemark}$ \\
\bottomrule
\end{tabular*}
\end{center}
\caption{
The tree-level four-particle quark-antiquark antennae and their distinct crossings.
$\widetilde{A}^0_4$
is symmetric under the interchange of the two photon-like gluons.
The nonidentical-flavour antenna $B^0_4$ is separately symmetric under interchange of $\qp$ with $\qbp$ and of $\q$ with $\qb$. \label{tab:qq}
}
\begin{center}
\begin{tabular*}{0.9 \textwidth} { l @{$\qquad$} l }
\toprule
\multicolumn{2}{l}{quark-gluon antennae}\\
\midrule
$D^0_4$	&
$D^0_4 \big( \cross{ \q },	\cross{ \gluon }, 	\gluon,			\gluon 			\big)$,
$D^0_4 \big( \cross{ \q },	\gluon,	 		\cross{ \gluon },		\gluon 			\big)$,
$D^0_4 \big( \q,		\cross{ \gluon },		\cross{ \gluon },		\gluon 			\big)$,
$D^0_4 \big( \q,		\cross{ \gluon },		\gluon,			\cross{ \gluon } 		\big)$
\\[3pt]
$E^0_4$	&
$E^0_4 \big( \cross{ \q },	\cross{ \qp },		\qbp,				\gluon			\big)$,
$E^0_4 \big( \cross{ \q },	\qp,				\qbp,				\cross{ g }			\big)$,
$E^0_4 \big( \q,		\cross{ \qp },	\cross{ \qbp },	\gluon			\big)$,
$E^0_4 \big( \q,		\cross{ \qp },	\qbp,			\cross{ g }		\big)$
\\[3pt]
$\widetilde{E}^0_4$	&
$\widetilde{E}^0_4 \big( \cross{ \q },	\cross{ \qp },		\qbp,				\gluon			\big)$,
$\widetilde{E}^0_4 \big( \cross{ \q },	\qp,				\qbp,				\cross{ g }			\big)$,
$\widetilde{E}^0_4 \big( \q,		\cross{ \qp },	\cross{ \qbp },	\gluon			\big)$,
$\widetilde{E}^0_4 \big( \q,		\cross{ \qp },	\qbp,			\cross{ g }		\big)$
\\[3pt]
\bottomrule
\end{tabular*}
\end{center}
\caption{The tree-level four-particle quark-gluon antennae and their distinct crossings.
Due to the cyclic colour connection, $D^0_4$ is symmetric under interchange of the second and fourth gluon. \label{tab:qg}
}
\begin{center}
\begin{tabular*}{0.9 \textwidth} { l @{$\qquad$} l }
\toprule
\multicolumn{2}{l}{gluon-gluon antennae}\\
\midrule
$F^0_4$	&
$F^0_4 \big( \cross{ \gluon },	\cross{ \gluon }, 	\gluon,			\gluon 			\big)$,
$F^0_4 \big( \cross{ \gluon },	\gluon,	 		\cross{ \gluon },		\gluon 			\big)$
\\[3pt]
$G^0_4$	&
$G^0_4 \big( \cross{ \gluon },			\q,				\qb,			\cross{ \gluon }		\big)$,
$G^0_4 \big( \cross{ \gluon },			\cross{ \q },		\qb,			\gluon			\big)$,
$G^0_4 \big( \cross{ \gluon },			\q,				\cross{ \qb },	\gluon			\big)$,
$G^0_4 \big( \gluon,				\cross{ \q },		\cross{ \qb },		\gluon			\big) $
\\[3pt]
$\widetilde{G}^0_4$	&
$\widetilde{G}^0_4 \big( \cross{ \gluon },			\q,				\qb,			\cross{\gluon}		\big)$,
$\widetilde{G}^0_4 \big( \cross{ \gluon },			\cross{ \q },		\qb,			\gluon			\big)$,
$\widetilde{G}^0_4 \big( \gluon,				\cross{ \q },		\cross{ \qb },	\gluon			\big) $
\\[3pt]
$H^0_4$ &
$H^0_4 \big(		\cross{ \q },		\cross{ \qb },		\qp,			\qbp			\big)$,
$H^0_4 \big(		\cross{ \q },		\qb,				\cross{ \qp },	\qbp			\big)$
\\[3pt]
\bottomrule
\end{tabular*}
\end{center}
\caption{
The tree-level four-particle gluon-gluon antennae and their distinct crossings.
$F^0_4$ is symmetric under cyclic interchange of its arguments.
$\widetilde{G}^0_4$ is symmetric under the interchange of the two photon-like 
gluons as well 
as under the interchange of $\q$ with $\qb$.
$H^0_4$ has three symmetries, $\q \leftrightarrow \qb$, $\qp \leftrightarrow \qbp$ 
and the flavour renaming $\q \leftrightarrow \qp$, $\qb \leftrightarrow \qbp$.
\label{tab:gg}}
\end{table}

Any two particles of a four-particle final-final antenna can be crossed 
to the initial state to obtain an initial-initial antenna; 
therefore one final-final four-particle antenna gives rise 
to six initial-initial antennae. 
Due to symmetries, at most four of these initial-initial antennae are different. 
The independent crossings~\cite{mathias}
 are listed in Tables~\ref{tab:qq}--\ref{tab:gg}.
To make the colour connection clear, in this list we write out the arguments 
of the antennae explicitly, i.e we write $X^0_4 \left( \cross{\imath}, j, \cross{k}, l \right)$
(where $\cross{ \imath }$ denotes an incoming particle) instead of $X^0_{ik,jl}$. 
The initial-initial antennae which are free of singular limits 
are not needed for the construction of subtraction terms, 
but their integrated form could prove useful for cross-checks. 

\subsection{Phase space factorisation and mappings}
The construction of subtraction terms requires a mapping from the original set
of momenta onto a reduced set. The mapping interpolates between the different
soft and collinear limits which the subtraction term regulates.
An appropriate mapping for the initial-initial case, both for single
and double unresolved configurations, has been discussed
in~\cite{hadant}. By requiring momentum conservation and phase space
factorisation, the phase space mapping is strongly constrained.
The remapping of initial state momenta can only be a rescaling, since
any transversal component would spoil the phase space factorisation.
For two unresolved partons $j$ and $k$, a complete factorisation of the phase
space into a convolution of an $m$-particle phase space
depending on redefined momenta only and the phase space of the unresolved 
partons $j$ and $k$ can be achieved with a Lorentz boost.
This boost maps the momentum $q \;=\; \pa + \pb - \fmom{j} - \fmom{k} $\,, with $q^2>0$\,
and $\pa, \pb$ being the momenta of the hard emitters,
into the momentum $\tilde{q} \,=\, \xahat \pa + \xbhat \pb $\,, where $\xahat$ and $\xbhat$
are fixed in terms of the invariants as follows:
\begin{equation}
 \xahat = \left(\frac{s_{12}-s_{j2}-s_{k2}}{s_{12}-s_{1j}-s_{1k}}
\;    \frac{q^2}{s_{12}}
     \right)^{\frac{1}{2}},
     \quad
     \xbhat =\left(\frac{s_{12}-s_{1j}-s_{1k}}{s_{12}-s_{j2}-s_{k2}}
\;    \frac{q^2}{s_{12}}
     \right)^{\frac{1}{2}}
\end{equation}
These two definitions guarantee the overall momentum conservation in the
mapped momenta and the correct soft and collinear behaviours\,.
The two momentum fractions $\xahat$ and $\xbhat$ satisfy the following limits in
double unresolved configurations:
\begin{enumerate}
\item $j$ and $k$ soft: $\xahat\rightarrow 1$, $\xbhat\rightarrow 1$,
\item $j$ soft and $k_{k}=z_1p_1$: $\xa\rightarrow 1-z_1$, $\xb\rightarrow 1$,
\item $k_{j}=z_1p_1$ and $k_{k}=z_2p_2$:
  $\xahat\rightarrow 1-z_1$, $\xbhat\rightarrow 1-z_2$,
\item $k_{j}=z_1p_1$, $\fmom{k} = z_2 p_1$: $\xahat\rightarrow 1-z_1-z_2$,
  $\xb\rightarrow 1$,
\end{enumerate}
and all the limits obtained from the ones above by the exchange of $p_1$ with
$p_2$ and of $k_j$ with $k_k$.
The construction of NNLO antenna subtraction terms also requires that
all single unresolved limits of the four-parton initial-initial antenna
functions $X_{il,jk}$, with radiators $i$ and $l$,
have to be subtracted, such that the resulting subtraction term is active only
in its double unresolved limits. A systematic subtraction of these single
unresolved limits through products of two three-parton antenna functions can be
performed only if the NNLO phase space mapping turns into an NLO phase space
mapping in its single unresolved limits. A detailed discussion of
the corresponding translation between these 
two momentum mappings can be found in~\cite{hadant}.

The factorisation of the $(m+2)$-parton phase space into 
an $m$-parton phase space and an antenna phase space involving the unresolved
partons $j$ and $k$ given in eq.~\eqref{eq:psii4} can equivalently
be written as
\begin{equation}
\begin{split}
\d\Phi_{m+2}( \psmomenta )&=
\d\Phi_{m}(\fmomentared;\xa \pa,\xb \pb)
\\
&\quad \times \; {\cal J} \;\delta(q^2-\xa\,\xb\,s_{12})\,
\delta(2\,(\xb \pb-\xa \pa )\cdot q)\,
\\
& \quad \times \;[\d k_j]\;[\d k_k]\;\d \xa\;\d \xb\,,
\label{PS}
\end{split}
\end{equation}
where ${\cal J}$ is the Jacobian factor defined by
\[
{\cal J}=s_{12}\,\left(\xa(s_{12}-s_{1j}-s_{1k}) + \xb
(s_{12}-s_{2j}-s_{2k})\right)\,.
\]

Using this  phase space parametrisation, we can express the 
integrated initial-initial antenna functions (\ref{eq:x4intii}) as:
\begin{equation}
\label{eq:intant}
{\cal X}^0_{il, jk}(x_1,x_2,\ep)=\frac{1}{[C(\epsilon)]^2}\int [{\rm d} k_j][{\rm d} k_k] \; {\cal J}\; x_1\; x_2\; \delta(C_1)\,\delta(C_2)\,X^0_{il,jk}\,,
\end{equation}
where
\begin{eqnarray}
\label{constraints}
  \nonumber
  C_1 &=& q^2 - \xa \, \xb  \, s_{12}\,, \\
  C_2 &=& 2 (\xb \pb - \xa \pa) \cdot q\,.
\end{eqnarray} 
In the following section, we describe how the integrals (\ref{eq:intant}) are carried out by means 
of a reduction to phase space master integrals with constraints (\ref{constraints}) and a
derivation of the master integrals from their differential equations.

\section{Integration over the double real radiation phase space}
\label{sec:int}

The initial-initial antenna functions have the scattering kinematics
\begin{displaymath}
\pa + \pb \to \fmom{j} + \fmom{k} + q \;,
\end{displaymath}
where $q$ is the momentum of the outgoing colourless particle.
The momenta satisfy:
\begin{displaymath}
\pa^2=\pb^2 =0, \;\;  \fmom{j}^2= \fmom{k}^2 =0, \;\;
q^2 = \tilde{q}^2 = x_1 \,x_2 \,s_{12}\,.
\end{displaymath}
The four-parton initial-initial antennae
need to be integrated over the phase space of the unresolved partons $j$ 
and $k$. This integration yields a result which depends only on $q^2$, 
$\xa$ and $\xb$. 
The dependence on $q^2$ is only multiplicative, 
according to the mass dimension of the integral.

Using the optical theorem, these phase space integrals are expressed as 
cuts of forward scattering two-loop diagrams, where the $\delta$-function 
conditions fixing $\xa$ and $\xb$ are introduced through non-standard 
propagators~\cite{dy,amhiggs}. 
The full set of propagators appearing in the four-parton antenna functions is:
\begin{eqnarray}
D_{j1} &=& (p_1-k_j)^2 \;,\nonumber \\
D_{k1} &=& (p_1-k_k)^2\;, \nonumber \\
D_{j2} &=& (p_2-k_j)^2\;, \nonumber \\
D_{k2} &=& (p_2-k_k)^2\;, \nonumber \\
D_{jk} &=& (k_j+k_k)^2\;, \nonumber \\
D_{jk1} &=& (p_1-k_j-k_k)^2\;, \nonumber \\
D_{jk2} &=& (p_2-k_j-k_k)^2\;, \nonumber \\
D_{j12} &=& (p_1+p_2-k_j)^2\;, \nonumber \\
D_{k12} &=& (p_1+p_2-k_k)^2\;, \nonumber \\
D_{j} &=& k_j^2\;, \nonumber \\
D_{k} &=& k_k^2\;, \nonumber \\
D_{jk12} &=& (p_1+p_2-k_j-k_k)^2 - q^2\;, \nonumber\\
D_{jk123} &=& (p_3 + p_1 + p_2 - k_j-k_k)^2 \;,
\end{eqnarray}
where $p_3 = x_2 \,p_2 -x_1 \,p_1$. The cut propagators are $D_{j}$, $D_{k}$, $D_{jk12}$,
$D_{jk123}$.
Only integrals in which all cut propagators appear in the denominator
are non-vanishing.
Besides the four cut propagators,
the phase space integrals of the antenna functions contain at most four further propagators.  
Any scalar product involving the integration momenta can be written as a linear combination 
of an appropriately chosen set of seven linear independent propagators (including the four 
cut propagators). Using this, we are able to express all integrands appearing in the phase 
space integration of the antenna functions in terms of a linearly independent set of seven 
of the above propagators. For each of these sets, we derived relations between different 
integrals based on integration-by-parts identities~\cite{ibp}, which are solved by using the 
Laporta algorithm, using the implementation in \cite{fire}. 
As a result, all phase space integrals can be expressed in terms 
of  20 independent phase space master integrals. 
Of these, 10 were already computed in~\cite{mathias}. 

The set of master integrals which we denote by ${ I}_i(x_1,x_2,\ep)$ 
are functions of $x_1$, $x_2$ and $\ep$.  
We begin by factoring out the leading behaviour of the master integrals
$I_i(x_1,x_2, \ep)$ in the limits $x_1 \to 1$
and $x_2 \to 1$, keeping the exact $\ep$-dependence:
\begin{equation}
{ I}_i(x_1,x_2,\ep) = (1-x_1)^{m_1-2 \,\ep} \;(1-x_2)^{m_2 - 2 \,\ep}\;
{\cal F}_i(x_1,x_2,\ep).
\end{equation}
The integers $m_1,m_2$ are characteristic to each master integral.
The functions ${\cal F}_i(x_1,x_2,\ep)$ are regular  at $x_1=1$,
at $x_2=1$, and at $x_1= x_2\, =1$
and can be calculated as Laurent series with, at most, second order
poles in $\ep$.

The integrated antennae given by ${\cal X}(x_1,x_2,\ep)$ are linear
combinations of these master integrals ${ I}_i(x_1,x_2,\ep)$, with
coefficients containing poles in $\ep$, as
well as in $(1-x_1)$ and $(1-x_2)$. After the masters have been inserted
into the integrated antennae, those take the form
%---------------
\begin{equation}
{\cal X}(x_1,x_2,\ep)
 =
(1-x_1)^{-1-2\,\ep}
(1-x_2)^{-1-2\,\ep}\;
{\cal R}(x_1,x_2,\ep),
\end{equation}
%---------------
where ${\cal R}(x_1,x_2,\ep)$ is regular at the boundaries $x_1 = 1$,
$x_2 = 1$, and at $x_1 = x_2 = 1$. The $\ep$-expansion of the
singular factors $(1-x_i)^{-1-2\ep}$ is done in the form
of distributions:
%---------------
\begin{equation}
\label{eq:plusdist}
(1-x_i)^{-1-2\ep}\,=\, -\frac{1}{2\ep} \, \delta(1-x_i) + \sum_{n}
\frac{(-2\ep)^{n}}{n!}\mathcal{D}_{n}(x_i)\,,
\end{equation}
with
\begin{equation}
\mathcal{D}_{n}(x_i)=\left(\frac{\ln^{n}(1-x_i)}{1-x_i}\right)_{+}\,.
\end{equation}
%---------------

To evaluate the integrated antennae,
we decompose the phase space into four regions depending on the values 
of $x_1$ and $x_2$. Those regions are given by:
\begin{itemize}
\item $x_1\,\neq\,1$, \,$x_2\,\neq\,1$, which we refer to as the hard region
\item $x_1\,=\,1$,\, $x_2\,\neq\,1$, and  $x_1\,\neq\,1$,\, $x_2\,=\,1$,
referred to as collinear regions
\item $x_1\,=\,1$,\, $x_2\,=\,1$, which we denote the soft region\,.
\end{itemize}
In the hard region ($x_1 \neq 1,\, x_2 \neq 1$), 
harmonic polylogarithms of weight two appear 
in the ${\cal O}(\ep^0)$ term of ${\cal R}$. Therefore,
the $\ep$-expansion of the master integrals in the hard region is needed 
at least up to the order at which terms of 
transcendentality
two appear.

In the collinear regions ($x_1 = 1$ or $x_2 = 1$), 
since the expansion in distributions in eq.~(\ref{eq:plusdist}) generates  
additional $1/\ep$ factors, the function ${\cal R}$ is required 
up to ${\cal O}(\ep)$, where harmonic polylogarithms of weight 3 appear.
The masters evaluated in the collinear region therefore need  to be expanded 
at least up to this order. 

Finally, in the soft region ($x_1=x_2=1$),
since the expansion of the distributions in eq.~(\ref{eq:plusdist}) generates
additional $1/\ep^2$ coefficients, the function ${\cal R}$ is required 
up to ${\cal O}(\ep^2)$ where transcendental constants of weight 4 appear. 
In this region, we evaluated all master integrals by direct integration. These 
integrations closely resemble the integrals appearing in the two-loop soft function, and 
we used the same techniques as were applied in~\cite{monni}, expanding the 
resulting hypergeometric functions in $\e$ using {\tt HypExp}~\cite{hypexp}. 

The master integrals in both collinear regions and in the hard region were determined 
from their differential equations in $\xa$ and $\xb$, using the results from the soft region 
as boundary condition. The results for all ten new master integrals
can be expressed in terms of harmonic polylogarithms~\cite{hpl} and a few 
specific combinations of logarithms and polylogarithms.
They 
are listed in the appendix. 

Insertion of the master integrals and Laurent  expansion of the overall factors according to 
eq.~(\ref{eq:plusdist}) yields the integrated antenna functions ${\cal X}(x_1,x_2,\ep)$.

\section{Results}
\label{sec:results}
The initial-initial antenna functions are listed in 
Tables~\ref{tab:qq}--\ref{tab:gg}. 
The full results for their integrated forms are too lengthy to be
quoted here, a separate FORM file containing those is attached 
with the arXiv submission of this paper.
In this section, we give at most two terms of their Laurent 
expansion around $\e=0$ below.
From each antenna function, we have omitted a common factor $(q^2)^{-2\e}$. This factor differs from the 
convention used in~\cite{mathias}, where a common factor  $(s_{12})^{-2\e}$
had been taken out. It is however in line with the final-final antenna functions derived 
in~\cite{ourant} and the initial-final antenna functions derived in~\cite{gionata}, which are all 
normalised to  $(q^2)^{-2\e}$. 

The leading pole structure of those antenna functions containing a singularity from 
two soft gluons or from a soft quark-antiquark pair can be predicted from infrared factorisation, 
and is universal between the final-final, initial-final and initial-initial configurations (in the former 
two cases, a single antenna function sometimes contains multiple soft gluon limits, which are 
then additive). For two colour-unconnected (abelian) soft gluons, we expect a leading pole of 
$1/\e^4$, for two colour-connected (non-abelian)
soft gluons a leading pole of $3/4/\e^4$ and for a soft 
quark-antiquark pair a soft pole of $-1/12/\e^3$.  A similar pattern can be seen in the 
integrated one-loop antenna functions~\cite{ourant,gionata,monni2}, where the 
difference between abelian and non-abelian gluonic contributions can be explained by 
the soft-gluon current at the one-loop order~\cite{cg}. The one-loop soft-gluon current 
contribution cancels between the double real radiation antennae and the one-loop 
antennae~\cite{our2j,ourant}. 

\subsection{Quark-antiquark antenna functions}
The leading poles of the 
integrated initial-initial antenna functions arising from crossings of the quark-antiquark antennae are as follows:
\begin{eqnarray}
{\cal A}_{12} &\equiv& 
{\cal A}^0_4 \big( \cross{ \q }, \gluon , \gluon,\cross{\qb}\big) \nonumber \\ &=& 
\hphantom{+}\frac{1}{\e^4}\frac{3}{4} \delta (1-x_1) \delta (1-x_2)
\nonumber\\&&
+\frac{1}{\e^3}\left[
\frac{11}{24} \delta(1-x_1) \delta(1-x_2)+
\frac{3}{4} \delta (1-x_2) \big( 1+ x_1 -2 \mathcal{D}_0(x_1) \big)
\right. \nonumber\\&& \hphantom{+\frac{1}{\e^4}} \, \left.
+\frac{3}{4} \delta (1-x_1) \big( 1 + x_2 -2 \mathcal{D}_0(x_2) \big)
\right]
+ \mathcal{O} \left( \e^{-2} \right)
\,,  \\
{\cal A}_{13} &\equiv& 
{\cal A}^0_4 \big( \cross{ \q }, \cross{\gluon} , \gluon,\qb\big) \nonumber \\ &=&  
\hphantom{+}\frac{1}{\e^3}\frac{1}{4} \left(1-2 x_2 + 2 x_2^2\right) \delta (1-x_1)
+ \mathcal{O} \left( \e^{-2} \right)
\,, \\
{\cal A}_{14} &\equiv& 
{\cal A}^0_4 \big( \cross{ \q }, \gluon , \cross{\gluon},\qb\big) \nonumber \\ &=& 
\hphantom{+}\frac{1}{\e^3}\frac{1}{2} \left(1-2 x_2+2 x_2^2\right) \delta (1-x_1)
+ \mathcal{O} \left( \e^{-2} \right)
\,,\\
{\cal A}_{34} &\equiv& 
{\cal A}^0_4 \big( \q , \cross{\gluon} , \cross{\gluon},\qb\big) \nonumber \\ &=&  
\hphantom{+}\frac{1}{\e^2}\frac{1}{4} \left( 1 - 2 x_1 + 2 x_1^2 \right) \left( 1 - 2 x_2 + 2 x_2^2 \right)
+ \mathcal{O} \left( \e^{-1} \right)
\,,\\
\tilde{{\cal A}}_{12} &\equiv& 
\tilde{{\cal A}}^0_4 \big( \cross{\q} , {\gluon} , {\gluon},\cross{\qb}\big) \nonumber \\ &=& 
\hphantom{+}\frac{1}{\e^4}\delta (1-x_1) \delta (1-x_2)
\nonumber\\&&
+\frac{1}{\e^3}\left[\delta (1-x_2) \left(1+ x_1 -2 \mathcal{D}_0(x_1)\right)+\delta (1-x_1) \left(1+x_2-2 \mathcal{D}_0(x_2)\right)\right]
\nonumber\\&&
+ \mathcal{O} \left( \e^{-2} \right)
\,,\\
\tilde{{\cal A}}_{13} &\equiv& 
\tilde{{\cal A}}^0_4 \big( \cross{\q} , \cross{\gluon} , \gluon,\qb\big) \nonumber \\ &=& 
\hphantom{+}\frac{1}{\e^3} \frac{1}{2} \left(1-2 x_2+2 x_2^2\right) \delta (1-x_1)
+ \mathcal{O} \left( \e^{-2} \right)
\,,\\
\tilde{{\cal A}}_{34} &\equiv& 
\tilde{{\cal A}}^0_4 \big( \q , \cross{\gluon} , \cross{\gluon},\qb\big) \nonumber \\ &=& 
\hphantom{+}\frac{1}{\e^2}\frac{1}{2} \left(1-2 x_1+2 x_1^2\right) \left(1-2 x_2+2 x_2^2\right)
+ \mathcal{O} \left( \e^{-1} \right)
\,,\\
{\cal B}_{12} &\equiv& 
{\cal B}^0_4 \big( \cross{ \q }, \qp , \qbp,\cross{\qb}\big) \nonumber \\ &=& 
\hphantom{+}\frac{1}{\e^3}\left[-\frac{1}{12} \delta (1-x_1) \delta (1-x_2)\right]
\nonumber\\&&
+\frac{1}{\e^2}\left[
-\frac{5}{36} \delta (1-x_1) \delta(1-x_2)
-\frac{1}{12} \delta (1-x_2) \left( 1 + x_1 - 2 \mathcal{D}_0(x_1) \right)
\right. \nonumber \\&& \hphantom{+\frac{1}{\e^2}} \, \left.
-\frac{1}{12} \delta (1-x_1) \left( 1 + x_2 - 2 \mathcal{D}_0(x_2) \right)
\right]
+ \mathcal{O} \left( \e^{-1} \right)
\,,\\
{\cal B}_{13} &\equiv& 
{\cal B}^0_4 \big( \cross{ \q }, \cross{\qp} , \qbp,{\qb}\big) \nonumber \\ &=& 
\hphantom{+}\frac{1}{\e^2}\delta (1-x_1) \left(\frac{1}{4} \left(x_2+1\right) H(0;x_2)+\frac{\left(1-x_2\right) \left(4 x_2^2+7 x_2+4\right)}{24 x_2}\right)
\nonumber\\&&
+ \mathcal{O} \left( \e^{-1} \right)
\,,\\
{\cal B}_{34} &\equiv& 
{\cal B}^0_4 \big( { \q },\cross{ \qp} , \cross{\qbp},\cross{\qb}\big) \nonumber \\ &=&  
\mathcal{O} \left( \e^0 \right)
\,,\\
{\cal C}_{12} &\equiv& 
{\cal C}^0_4 \big( \cross{ \q }, {\qb}  , \q,\cross{\qb} \big) \nonumber \\ &=& 
\hphantom{+}\frac{1}{\e}\Bigg[
\frac{\delta (1-x_1)}{{48 \left(1-x_2\right)}} 
\left(
	(1+x_2^2) \left[ 6 H(0,0;x_2) +6  H(1,0;x_2) + \pi^2 \right]
	\right. \nonumber \\&& \left. \hphantom{\frac{1}{\e} \Bigg[ \frac{\delta (1-x_1)}{{48 \left(1-x_2\right)}}  } \quad
	+ 3 ( 5 - 2 x_2^2) H(0;x_2)
	+ 3 (1-x_2) (8-7 x_2)
\right)
\Bigg]
\nonumber \\&&
+ \mathcal{O} \left( \e^{0} \right)
\,,\\
{\cal C}_{13} &\equiv& 
{\cal C}^0_4 \big( \cross{ \q }, {\qb} , \cross{\q},{\qb}\big) \nonumber \\ &=& 
\hphantom{+}\frac{1}{\e}\Bigg[
\frac{\delta (1-x_1)}{{24 \left(1+ x_2\right)}} 
\left(
	(1+x_2^2) \left[ 12 H(-1,0;x_2) - 6  H(0,0;x_2)  + \pi^2 \right]
	\right. \nonumber \\&& \left. \hphantom{\frac{1}{\e} \Bigg[ \frac{\delta (1-x_1)}{{48 \left(1-x_2\right)}}  } \quad
	- 12 (1-x_2^2)  -6 (1+x_2)^2 H(0;x_2)
\right)
\Bigg]
+ \mathcal{O} \left( \e^{0} \right)
\,,\\
{\cal C}_{23} &\equiv& 
{\cal C}^0_4 \big( { \q }, \cross{\qb} , \cross{\q},{\qb}\big) \nonumber \\ &=& 
\mathcal{O} \left( \e^0 \right)
\,,\\
{\cal C}_{24} &\equiv& 
{\cal C}^0_4 \big( { \q }, {\qb} , \cross{\q},\cross{{\qb}}\big) \nonumber \\ &=&  
\mathcal{O} \left( \e^0 \right)
\,.
\end{eqnarray}
We observe that ${\cal B}_{34}$, ${\cal C}_{23}$ and ${\cal C}_{34}$ are finite, as expected. Moreover, the leading pole structure of 
${\cal A}_{12}$ (non-abelian), $\tilde{{\cal A}}_{12}$ (abelian)
and ${\cal B}_{12}$ is as predicted from the 
double soft factorisation. ${\cal B}_{12}$, ${\cal B}_{13}$ and ${\cal B}_{34}$ were computed 
previously in \cite{mathias}.

\subsection{Quark-gluon antenna functions}
The leading poles of the integrated initial-initial antenna functions obtained from crossing quark-gluon antennae read:
\begin{eqnarray}
{\cal D}_{12} &\equiv&  
{\cal D}^0_4 \big( \cross{ \q }, \cross{\gluon} , \gluon,\gluon\big) \nonumber \\ &=& 
\hphantom{+}\frac{1}{\e^4}\frac{3}{4} \delta (1-x_1) \delta (1-x_2)
\nonumber\\&&
+\frac{1}{\e^3}\left[
\frac{11}{24} \delta(1-x_1) \delta (1-x_2) 
+ \frac{3}{4}  \delta(1-x_2) \left( 1+ x_1 -2 \mathcal{D}_0(x_1)\right)
\right. \nonumber\\&& \hphantom{+\frac{1}{\e^4}} \, \quad \left.
+\frac{3}{2}  \delta (1-x_1) \left( x_2^2-x_2-\frac{1}{x_2}+2  -\mathcal{D}_0(x_2) \right)
\right]
+ \mathcal{O} \left( \e^{-2} \right)
\,, \\
{\cal D}_{13} &\equiv& 
{\cal D}^0_4 \big( \cross{ \q }, {\gluon} , \cross{\gluon},\gluon\big) \nonumber \\ &=& 
\hphantom{+}\frac{1}{\e^4}\delta (1-x_1) \delta (1-x_2)
\nonumber\\&&
+\frac{1}{\e^3}\Bigg[
\delta (1-x_2) \left( 1+ x_1 -2 \mathcal{D}_0(x_1)\right)
 \nonumber\\&& \hphantom{+\frac{1}{\e^4}} \, 
+2 \delta (1-x_1) \left(  x_2^2-x_2-\frac{1}{x_2}+2 - \mathcal{D}_0(x_2)\right)
\Bigg]
+ \mathcal{O} \left( \e^{-2} \right)
\,, \\
{\cal D}_{23} &\equiv& 
{\cal D}^0_4 \big( { \q }, \cross{\gluon} , \cross{\gluon},\gluon\big) \nonumber \\ &=& 
\hphantom{+}\frac{1}{\e^3}\left[\frac{1}{4} \left(1-2 x_2+2 x_2^2\right) \delta (1-x_1)+\frac{1}{2} \left(1-2 x_1 + 2 x_1^2\right) \delta (1-x_2)\right]
\nonumber\\&&
+ \mathcal{O} \left( \e^{-2} \right)
\,,\\
{\cal D}_{24} &\equiv& 
{\cal D}^0_4 \big( { \q }, \cross{\gluon} , \gluon,\cross{\gluon}\big) \nonumber \\ &=&  
\hphantom{+}\frac{1}{\e^3}\left[\frac{1}{2} \left(1-2 x_2+2 x_2^2\right) \delta (1-x_1)+\frac{1}{2} \left(1-2 x_1+2 x_1^2\right) \delta (1-x_2)\right]
\nonumber\\&&
+ \mathcal{O} \left( \e^{-2} \right)
\,,\\
{\cal E}_{12} &\equiv&  
{\cal E}^0_4 \big( \cross{ \q }, \cross{\qp} , \qbp,\gluon\big) \nonumber \\ &=& 
\hphantom{+}\frac{1}{\e^3}\frac{\left(2-2 x_2+ x_2^2\right) \delta (1-x_1)}{2 x_2}
+ \mathcal{O} \left( \e^{-2} \right)
\,,\\
{\cal E}_{14} &\equiv& 
{\cal E}^0_4 \big( \cross{ \q }, {\qp} , \qbp,\cross{\gluon}\big) \nonumber \\ &=&
\hphantom{+}\frac{1}{\e^3}\left[-\frac{1}{12} \delta (1-x_1) \delta (1-x_2)\right]
\nonumber\\&&
+\frac{1}{\e^2}\left[
-\frac{5}{36} \delta(1-x_1) \delta (1-x_2)
- \frac{1}{12} \delta (1-x_2) \left(1 + x_1 - 2 \mathcal{D}_0(x_1) \right)
 \right. \nonumber \\&& \left.  \hphantom{+ \frac{1}{\e^3}} \quad 
+\delta (1-x_1) \left(\frac{6 x_2 H(0;x_2)(1+x_2) - 8 x_2^3 + x_2^2 - 5 x_2 + 8}{24 x_2}
\right. \right. \nonumber \\&& \left. \left.  \hphantom{+ \frac{1}{\e^3} + \delta(1-x_1)} \qquad
+\frac{1}{6} \mathcal{D}_0(x_2)\right)
\right]
+ \mathcal{O} \left( \e^{-1} \right)
\,,\\
{\cal E}_{23} &\equiv& 
{\cal E}^0_4 \big( { \q }, \cross{\qp} , \cross{\qbp},\gluon\big) \nonumber \\ &=& 
\hphantom{+}\frac{1}{\e^2}\frac{x_1 x_2 \left(1+x_1 x_2\right)^2 \left(x_1^2+x_2^2-2\right)}{\left(x_1+x_2\right)^4}
+ \mathcal{O} \left( \e^{-1} \right)
\,,\\
{\cal E}_{24} &\equiv& 
{\cal E}^0_4 \big( { \q }, \cross{\qp} , \qbp,\cross{\gluon}\big) \nonumber \\ &=& 
\hphantom{+}\frac{1}{\e^2}\left[\delta (1-x_2) \left(
\frac{1}{4} \left(1+x_1\right) H(0;x_1)
+ \frac{\left(1-x_1\right) \left(4 x_1^2+7 x_1+4\right)}{24 x_1}\right)
\right. \nonumber \\&& \left. \hphantom{ + \frac{1}{\e^2} } \quad
+\frac{\left(2-2 x_1+x_1^2\right) \left(1-2 x_2+2 x_2^2\right)}{4 x_1}\right]
+ \mathcal{O} \left( \e^{-1} \right)
\,,\\
\tilde{{\cal E}}_{12} &\equiv&  
\tilde{{\cal E}}^0_4 \big( \cross{ \q }, \cross{\qp} , \qbp,\gluon\big) \nonumber \\ &=& 
\hphantom{+}\frac{1}{\e^3}\frac{\left(2-2 x_2+x_2^2\right) \delta (1-x_1)}{4 x_2}
+ \mathcal{O} \left( \e^{-2} \right)
\,,\\
\tilde{{\cal E}}_{14} &\equiv& 
\tilde{ {\cal E}}^0_4 \big( \cross{ \q }, {\qp} , \qbp,\cross{\gluon}\big) \nonumber \\ &=& 
\hphantom{+}\frac{1}{\e^2}\delta (1-x_1) \left(\frac{1}{2} \left(1+x_2\right) H(0;x_2)+\frac{\left(1-x_2\right) \left(4 x_2^2+7 x_2+4\right)}{12 x_2}\right)
\nonumber\\&&
+ \mathcal{O} \left( \e^{-1} \right)
\,,\\
\tilde{{\cal E}}_{23} &\equiv& 
\tilde{{\cal E}}^0_4 \big( { \q }, \cross{\qp} , \cross{\qbp},\gluon\big) \nonumber \\ &=& 
\hphantom{+}\frac{1}{\e^2}\frac{x_1 x_2 \left(1+x_1 x_2\right)^2 \left(x_1^2+x_2^2-2\right)}{\left(x_1+x_2\right)^4}
+ \mathcal{O} \left( \e^{-1} \right)
\,,\\
\tilde{{\cal E}}_{24} &\equiv& 
\tilde{{\cal E}}^0_4 \big( { \q }, \cross{\qp} , \qbp,\cross{\gluon}\big) \nonumber \\ &=& 
\hphantom{+}\frac{1}{\e}\Bigg[
\frac{\left(2 x_1^8+x_2 x_1^7-x_1^6+2 x_2^2 x_1^4-x_2 x_1^3-8 x_2^2 x_1^2+2 x_2 x_1+8 x_2^2\right) 
	 }{x_1^4} \cdot
\nonumber \\&& \hphantom{+ \frac{1}{\e} } \quad \quad
\cdot \big( G(-x_2;x_1) - G(-1;x_1) \big)
+\left(2 x_1^2+x_2 x_1-1\right) x_1^2 H(0;x_2)
\nonumber \\&& \hphantom{+ \frac{1}{\e} } \quad
+\frac{\left(1-x_2\right)}{6 \left(x_1+x_2\right)^3 x_1^3}
 \left(12 x_1^9+36 x_2 x_1^8-6 x_1^8+41 x_2^2 x_1^7  -19 x_2 x_1^7
\right. \nonumber \\&& \left. \hphantom{+ \frac{1}{\e} } \quad
-x_1^7+19 x_2^3 x_1^6-29 x_2^2 x_1^6-5 x_2 x_1^6+4 x_2^4 x_1^5-20 x_2^3 x_1^5-26 x_2^2 x_1^5
\right. \nonumber \\&& \left. \hphantom{+ \frac{1}{\e} } \quad
+15 x_2 x_1^5  -6 x_2^4 x_1^4-30 x_2^3 x_1^4+73 x_2^2 x_1^4+10 x_2 x_1^4-2 x_1^4-12 x_2^4 x_1^3
\right. \nonumber \\&& \left. \hphantom{+ \frac{1}{\e} } \quad
+86 x_2^3 x_1^3+50 x_2^2 x_1^3  -34 x_2 x_1^3+32 x_2^4 x_1^2+62 x_2^3 x_1^2-118 x_2^2 x_1^2
 \right. \nonumber \\&& \left. \hphantom{+ \frac{1}{\e} } \quad
+24 x_2^4 x_1-132 x_2^3 x_1-48 x_2^4
\right)
\Bigg]
\nonumber\\&&
+ \mathcal{O} \left( \e^{0} \right)
\,.
\end{eqnarray}
Again, the leading pole structure of 
${\cal D}_{12}$ (non-abelian), ${\cal D}_{13}$ (abelian) and ${\cal E}_{14}$ is as expected.
$\tilde{{\cal E}}_{12}$, $\tilde{{\cal E}}_{14}$, $\tilde{{\cal E}}_{23}$, $\tilde{{\cal E}}_{24}$ 
were computed already in  \cite{mathias}.

\subsection{Gluon-gluon antenna functions}
Finally, the leading poles of the integrated initial-initial gluon-gluon antenna functions are:
\begin{eqnarray}
{\cal F}_{12} &\equiv& 
{\cal F}^0_4 \big( \cross{ \gluon }, \cross{\gluon} , \gluon,\gluon\big) \nonumber \\ &=& 
\hphantom{+}\frac{1}{\e^4}\frac{3}{4} \delta (1-x_1) \delta (1-x_2)
\nonumber\\&&
+\frac{1}{\e^3}\left[
\frac{11}{24} \delta(1-x_1) \delta(1-x_2) 
+ \frac{3}{2} \delta (1-x_2) \ \left(x_1^2-x_1-\frac{1}{x_1}+2- \mathcal{D}_0(x_1)\right)
\right. \nonumber\\&& \hphantom{+\frac{1}{\e^4}} \, \quad \left.
+ \frac{3}{2} \delta (1-x_1) \ \left(x_2^2-x_2-\frac{1}{x_2}+2- \mathcal{D}_0(x_2)\right)
\right]
+ \mathcal{O} \left( \e^{-2} \right)
 \,,\\ 
{\cal F}_{13} &\equiv& 
{\cal F}^0_4 \big( \cross{ \gluon }, {\gluon} , \cross{\gluon},\gluon\big) \nonumber \\ &=& 
\hphantom{+}\frac{1}{\e^4}\delta (1-x_1) \delta (1-x_2)
\nonumber\\&&
+\frac{1}{\e^3}\left[
2 \delta (1-x_1) \ \left(x_2^2-x_2-\frac{1}{x_2}+2- \mathcal{D}_0(x_2)\right)
\right. \nonumber\\&& \hphantom{+\frac{1}{\e^4}} \,  \left.
+2\delta (1-x_2) \ \left(x_1^2-x_1-\frac{1}{x_1}+2- \mathcal{D}_0(x_1)\right)
\right]
+ \mathcal{O} \left( \e^{-2} \right)
\,, \\ 
{\cal G}_{12} &\equiv&  
{\cal G}^0_4 \big( \cross{ \gluon }, \q , \qb,\cross{\gluon}\big) \nonumber \\ &=& 
\hphantom{+}\frac{1}{\e^3}\left[-\frac{1}{12} \delta (1-x_1) \delta (1-x_2)\right]
\nonumber\\&&
+\frac{1}{\e^2}\left[
\frac{-5}{36} \delta(1-x_1) \delta(1-x_2)
\right. \nonumber \\&& \left. \hphantom{+ \frac{1}{\e} } \quad
+\delta (1-x_1) \left(\frac{6 x_2 H(0;x_2)(1+x_2) - 8 x_2^3 + x_2^2 - 5 x_2 + 8}{24 x_2}
\right. \right. \nonumber \\&& \left. \left. \hphantom{+ \frac{1}{\e} + \delta(1-x_1)} \qquad
+\frac{1}{6} \mathcal{D}_0(x_2)\right)
\right. \nonumber \\&& \left. \hphantom{+ \frac{1}{\e} } \quad
+\delta (1-x_2) \left(\frac{6 x_1 H(0;x_1)(1+x_1) - 8 x_1^3 + x_1^2 - 5 x_1 + 8}{24 x_1}
\right. \right. \nonumber \\&& \left. \left. \hphantom{+ \frac{1}{\e} + \delta(1-x_1)} \qquad
+\frac{1}{6} \mathcal{D}_0(x_1)\right)
\right]
+ \mathcal{O} \left( \e^{-1} \right)
\,, \\ 
{\cal G}_{13} &\equiv& 
{\cal G}^0_4 \big( \cross{ \gluon }, \cross{\q} , \qb,{\gluon}\big) \nonumber \\ &=& 
\hphantom{+}\frac{1}{\e^3}\frac{\left(2-2 x_2+x_2^2\right) \delta (1-x_1)}{2 x_2}
+ \mathcal{O} \left( \e^{-2} \right)
 \,,\\  
{\cal G}_{14} &\equiv& 
{\cal G}^0_4 \big( \cross{ \gluon }, \q , \cross{\qb},{\gluon}\big) \nonumber \\ &=&  
\hphantom{+}\frac{1}{\e^3}\frac{\left(2-2 x_2+x_2^2\right) \delta (1-x_1)}{2 x_2}
+ \mathcal{O} \left( \e^{-2} \right)
\,, \\  
{\cal G}_{34} &\equiv& 
{\cal G}^0_4 \big( { \gluon }, \cross{\q} , \cross{\qb},{\gluon}\big) \nonumber \\ &=& 
\hphantom{+}\frac{1}{\e^2}\frac{2 \left(1 + x_1 x_2\right) \left(x_2^2 x_1^4+x_2^4 x_1^2-4 x_2^2 x_1^2+x_1^2+x_2^2\right)}{\left(x_1+x_2\right)^4}
+ \mathcal{O} \left( \e^{-1} \right)
\,, \\  
\tilde{{\cal G}}_{12} &\equiv&  
\tilde{{\cal G}}^0_4 \big( \cross{ \gluon }, \q , \qb,\cross{\gluon}\big) \nonumber \\ &=& 
\hphantom{+}\frac{1}{\e^2}\left[
\delta (1-x_2) \left(\frac{1}{2} \left(1+x_1\right) H(0;x_1)+\frac{\left(1-x_1\right) \left(4 x_1^2+7 x_1+4\right)}{12 x_1}\right)
\right. \nonumber \\&& \left. \hphantom{ + \frac{1}{\e^2} } \quad
+\delta (1-x_1) \left(\frac{1}{2} \left(1+x_2\right) H(0;x_2)+\frac{\left(1-x_2\right) \left(4 x_2^2+7 x_2+4\right)}{12 x_2}\right)
\right]
 \nonumber \\&& 
+ \mathcal{O} \left( \e^{-1} \right)
\,, \\  
\tilde{{\cal G}}_{13} &\equiv& 
\tilde{{\cal G}}^0_4 \big( \cross{ \gluon }, \cross{\q} , \qb,{\gluon}\big) \nonumber \\ &=& 
\hphantom{+}\frac{1}{\e^3}\frac{\left(2-2 x_2+x_2^2\right) \delta (1-x_1)}{4 x_2}
+ \mathcal{O} \left( \e^{-2} \right)
\,, \\  
\tilde{{\cal G}}_{34} &\equiv& 
\tilde{{\cal G}}^0_4 \big( { \gluon }, \cross{\q} , \cross{\qb},{\gluon}\big) \nonumber \\ &=& 
\hphantom{+}\frac{1}{\e^2}\frac{2 \left(1+x_1 x_2\right) \left(x_2^2 x_1^4+x_2^4 x_1^2-4 x_2^2 x_1^2+x_1^2+x_2^2\right)}{\left(x_1+x_2\right)^4}
+ \mathcal{O} \left( \e^{-1} \right)
\,, \\  
{\cal H}_{12} &\equiv&  
{\cal H}^0_4 \big( \cross{ \q }, \cross{\qb} , \qp,\qbp\big) \nonumber \\ &=& 
\hphantom{+}\frac{1}{\e}\left[-\frac{\left(1+x_1 x_2\right) \left(x_2^2 x_1^4+x_2^4 x_1^2-4 x_2^2 x_1^2+x_1^2+x_2^2\right)}{3 \left(x_1+x_2\right)^4}\right]
+ \mathcal{O} \left( \e^{0} \right)
 \,,\\  
{\cal H}_{13} &\equiv& 
{\cal H}^0_4 \big( \cross{ \q }, {\qb} , \cross{\qp},\qbp\big) \nonumber \\ &=& 
\hphantom{+}\frac{1}{\e^2}\frac{\left(2-2 x_1+x_1^2\right) \left(2-2 x_2+x_2^2\right)}{4 x_1 x_2}
+ \mathcal{O} \left( \e^{-1} \right)
 \,.
\end{eqnarray}
 The leading pole structure of 
${\cal F}_{12}$ (non-abelian), ${\cal F}_{13}$ (abelian)
and ${\cal G}_{12}$ is as predicted from the 
double soft factorisation.
${\cal H}_{12}$ and ${\cal H}_{13}$
were computed already in  \cite{mathias}.

\section{Conclusions}
\label{sec:conc}
In this paper, we have completed the integration of double 
real radiation antenna functions in the kinematical situation where both 
radiator partons are in the initial state. A subset of our results, restricted to antenna functions 
containing a secondary fermion pair, was obtained earlier in~\cite{mathias}. With the results 
presented in this paper, combined with the final-final~\cite{ourant}, 
initial-final~\cite{gionata} and virtual one-loop initial-initial antennae~\cite{monni2}, 
all antenna functions required for the calculation of jet cross 
sections at hadron colliders are now available in unintegrated and integrated form. The 
implementation of the corresponding antenna subtraction terms 
into a numerical parton-level event generator program is thus readily feasible. 
 
Using the general method completed in this paper, NNLO QCD calculations for exclusive 
jet observables at hadron colliders are now in reach. The principal requirement for these 
calculations is the availability of the relevant two-loop matrix elements. Potential 
applications, where two-loop matrix elements have already been derived, are:
two-jet production~\cite{m2}, vector-boson-plus-jet 
production~\cite{3jme} and Higgs-boson-plus-jet production~\cite{hjme}.
Important steps have already been taken in the calculation of two-jet production at NNLO, where 
the unintegrated subtraction terms for purely 
gluonic processes have already been derived and tested for the double 
real radiation at tree-level~\cite{joao} and the single real radiation at one-loop~\cite{joaonew},
and in the incorporation of processes with quarks into this calculation~\cite{currie}. 
Finally, we notice that the integrated initial-initial antenna functions $\mathcal{F}_{12}$ and 
$\mathcal{F}_{13}$ derived here 
appear in the integrated subtraction terms of the all-gluon contribution to two-jet production.  
They contribute to the cancellation of all infrared poles when combined
 with all 
other integrated subtraction terms and with the virtual two-loop matrix 
elements~\cite{joaovv}. This pole cancellation has been verified~\cite{joaovv} and can be
regarded as a strong check on the correctness of our results presented
here for these integrated antennae $\mathcal{F}_{12}$ and $\mathcal{F}_{13}$.

\section*{Acknowledgements}
We would like to thank Nigel Glover and Joao Pires 
for many stimulating discussions.
Part of this work was done while AG and TG attended the 
``Physics with first data from the LHC'' program at the  Kavli Institute for 
Theoretical Physics (KITP) at UC Santa Barbara. 
This research is supported in part by
the Swiss National Science Foundation (SNF) under contracts
PP00P2-139192 and 200020-138206, and by 
the Research Executive Agency (REA) of the European Union under the Grant 
Agreement number PITN-GA-2010-264564 (LHCPhenoNet). 
The work of MR was supported under European Research 
Council Advanced Investigator Grant ERC--AdG--228301.

\appendix
\section{Master integrals}
\label{app:A}
The phase space integrals appearing in the integration of the initial-initial four-parton antenna 
functions can be reduced to a set of 20 independent master integrals, as described in Section~\ref{sec:int}. 
Of these integrals, the following 10  
were derived previously in~\cite{mathias}:
\begin{eqnarray}
% {1, SS[kb, p1]},
I_{1} & = & \int \dmom{j} \dmom{k} \delta(C_1) \delta(C_2)
        ( \fmom{k} \cdot p_1 )\,,
\nonumber\\\nonumber
% {2, 1},
I_{2} & = & \int \dmom{j} \dmom{k} \delta(C_1) \delta(C_2) \,,
\\\nonumber
% {3, Db12 SS[ka, p1] SS[kb - p1 - p2, p3]},
I_{3} & = & \int \dmom{j} \dmom{k} \delta(C_1) \delta(C_2)
      \frac{  - (\fmom{j} \cdot p_1) (\fmom{j} \cdot p_3)}{D_{k12}} \,,
\\\nonumber
% {4, Db12 SS[ka, p1]},
I_{4} & = & \int \dmom{j} \dmom{k} \delta(C_1) \delta(C_2)
      \frac{  (\fmom{j} \cdot p_1) }{D_{k12}}\,,
\\\nonumber
% {5, Db12 SS[kb - p1 - p2, p3]},
I_{5} & = & \int \dmom{j} \dmom{k} \delta(C_1) \delta(C_2)
      \frac{ - (\fmom{j} \cdot p_3) }{D_{k12}} \,,
\\\nonumber
% {6, Db12},
I_{6} & = & \int \dmom{j} \dmom{k} \delta(C_1) \delta(C_2)
      \frac{ 1 }{D_{k12}} \,,
\\\nonumber
% {7, Dab2},
I_{7} & = & \int \dmom{j} \dmom{k} \delta(C_1) \delta(C_2)
      \frac{ 1 }{D_{jk2}} \,,
\\\nonumber
% {8, Db2},
I_{8} & = & \int \dmom{j} \dmom{k} \delta(C_1) \delta(C_2)
      \frac{ 1 }{D_{k2}} \,,
\\\nonumber
% {14, Da12 Dab2},
I_{14} & = & \int \dmom{j} \dmom{k} \delta(C_1) \delta(C_2)
      \frac{ 1 }{D_{j12} D_{jk2}}\,,
\\
% {15, Da1 Db2}
I_{15} & = & \int \dmom{j} \dmom{k} \delta(C_1) \delta(C_2)
      \frac{ 1 }{D_{j1} D_{k2}} \,.
\end{eqnarray}

The remaining 10 integrals are:
\begin{eqnarray}
I_{13} &=& \int [\d k_j] \,[\d k_k] \, \delta (C_1) \delta (C_2) \,\frac{1}{D_{k2} D_{j12}}\,,\nonumber \\
I_{16} &=&  \int [\d k_j] \,[\d k_k] \, \delta (C_1) \delta (C_2) 
\,\frac{1}{D_{jk}\,  D_{j1}\, D_{j12}}\,,\nonumber \\
I_{17} &=& \int [\d k_j] \,[\d k_k] \, \delta (C_1) \delta (C_2) 
\,\frac{1}{D_{j1}\,  D_{k2}\, D_{jk2}}\,,\nonumber \\
I_{22} &=& \int [\d k_j] \,[\d k_k] \, \delta (C_1) \delta (C_2) 
\,\frac{1}{D_{j1}\,  D_{k2}\, D_{j12}}\,,\nonumber \\
I_{23} &=& \int [\d k_j] \,[\d k_k] \, \delta (C_1) \delta (C_2) 
\,\frac{1}{D_{k2}\,  D_{jk2}\, D_{j12}}\,,\nonumber \\
I_{24} &=&\int [\d k_j] \,[\d k_k] \, \delta (C_1) \delta (C_2) 
\,\frac{1}{D_{j1}\,  D_{jk2}\, D_{j12}} \,,\nonumber \\
I_{25} &=& \int [\d k_j] \,[\d k_k] \, \delta (C_1) \delta (C_2) 
\,\frac{1}{D_{j2}\,  D_{k2}\, D_{k12}}\,,\nonumber \\
I_{27} &=& \int [\d k_j] \,[\d k_k] \, \delta (C_1) \delta (C_2) 
\,\frac{1}{D_{j1}\,  D_{j2}\, D_{jk2}}\,,\nonumber \\
I_{28} &=& \int [\d k_j] \,[\d k_k] \, \delta (C_1) \delta (C_2) 
\,\frac{1}{D_{j1}\,  D_{j2}\, D_{k2}}\,,\nonumber \\
I_{29} &=& \int [\d k_j] \,[\d k_k] \, \delta (C_1) \delta (C_2) 
\,\frac{1}{D_{j2}\,  D_{jk}\, D_{k1}}\,.
\end{eqnarray}
Note that, the numeration of the integrals is not consecutive, since masters
related by exchange of $\xa$ and $\xb$ are listed only for one
crossing in both sets. 

In the new set, the integrals $I_{13}$, $I_{22}$, $I_{23}$ and
$I_{25}$ appear always in particular combinations, 
which can be made explicit by introducing new master integrals $M_{22}$, $M_{23}$ and
$M_{25}$ by:
\begin{eqnarray}
I_{22} &=& M_{22} + \frac{1}{2 x_1 x_2 s_{12}} I_{13}\,, \nonumber \\
I_{23} &=& M_{23} + \frac{1-2x_1^2 - 2x_1x_2}{2x_2(1-x_1^2)(x_1+x_2)s_{12}} I_{13}\,, \nonumber \\
I_{25} &=& M_{25} + \frac{x_1(x_1+2x_2)}{2x_2(1-x_1^2)s_{12}}I_{13}\,.
\end{eqnarray}
After expressing the 
integrated antenna functions in 
 this new basis it turns out that the remaining coefficients of $I_{13}$ are all proportional to 
 $\e$, such that $I_{13}$ is required only to lower order in $\e$ than the other 
 master integrals. 

\subsection{Hard region}
In the hard region, the new master integrals are:
\begin{eqnarray}
I_{13} &= & (s_{12})^{-2-2\e} \, S_\Gamma \, \frac{(1-x_1)^{-2\e}(1-x_2)^{-2\e}}{x_1} 
\frac{1}{\e} g_{13}(x_1,x_2)+ {\cal O}(\e^0) \;,
\\
I_{16} & = & (s_{12})^{-3-2\e} \, S_\Gamma \, \frac{(1-x_1)^{-2\e}(1-x_2)^{-2\e}}
{x_1 (1-x_2^2)} \Bigg[ \nonumber \\
& & - \frac{1}{\e^2} + \frac{1}{\e} \big( 
          - 3 G( - x_2;x_1)
          + G(-1;x_1)
          + 2 H(-1;x_2)
          - 3 H(0;x_2)
          \big) \nonumber \\
&&       + \frac{17\pi^2}{12}
          - 2 G(1/x_2;x_1) H(0;x_2)
          - 2 G(1/x_2,0;x_1)
          - 6 G( - x_2, - x_2;x_1)
\nonumber \\
&&            + 3 G( - x_2;x_1) H(-1;x_2)
          - 6 G( - x_2;x_1)H(0;x_2)
          + 3 G( - x_2,-1;x_1)
\nonumber \\
&&            + 3 G(-1, - x_2;x_1)
          + G(-1;x_1)  \log 2
          - 2 G(-1;x_1) H(-1;x_2)
          + 3 G(-1;x_1) H(0;x_2)
\nonumber \\
&&            - 2 G(-1,-1;x_1)
          + 2 G(0;x_1) H(0;x_2)
          + 4 G(0,0;x_1)
          - 3 G(1, - x_2;x_1)
\nonumber \\
&&            - G(1;x_1)  \log 2
          + 3 G(1;x_1) H(-1;x_2)
          - 3 G(1;x_1) H(0;x_2)
          + G(1,-1;x_1)
\nonumber \\
&&            - 2 G(1,0;x_1)
          - 4 H(-1,-1;x_2)
          + 6 H(-1,0;x_2)
          + 3 H(0,-1;x_2)
\nonumber \\
&&            - 4 H(0,0;x_2)
 \Bigg] + {\cal O} (\e)\;,
\\
I_{17} & = & (s_{12})^{-3-2\e} \, S_\Gamma \, \frac{(1-x_1)^{-2\e}(1-x_2)^{-2\e}}{x_1x_2^2 (1-x_1)} \Bigg[ 
-\frac{1}{2\e^2} \nonumber \\ &&+ \frac{1}{\e} \left( 
            \log 2
          - G(-1;x_1)
          + G(0;x_1)
          + H(0;x_2)
\right)\nonumber \\
&&           - 2  \log^2 2
          - \frac{\pi^2}{2}
          + G(1/x_2;x_1) H(-1;x_2)
          - G(1/x_2,-1;x_1)
          + G(1/x_2,0;x_1)\nonumber \\
&&
          - 3 G(-1, - x_2;x_1)
          + 2 G(-1;x_1)  \log 2
          - G(-1;x_1) H(-1;x_2)
          - G(-1;x_1) H(0;x_2)\nonumber \\
&&
          + 2 G(-1,-1;x_1)
          + G(-1,0;x_1)
          - 2 G(0;x_1)  \log 2
          - 2 G(0;x_1) H(0;x_2)\nonumber \\
&&
          + 2 G(0,-1;x_1)
          - 2 G(0,0;x_1)
          - 2 H(0;x_2)  \log 2
          - 2 H(0,0;x_2)
          - 2 H(1;x_2)  \log 2\nonumber \\
&&
          + 2 H(1,-1;x_2)
          - 4 H(1,0;x_2) 
           \Bigg] \nonumber \\
&& +(s_{12})^{-3-2\e} \, S_\Gamma \, \frac{(1-x_1)^{-2\e}(1-x_2)^{-2\e}}{x_1x_2^2 (1+x_1)} \Bigg[ 
\frac{1}{\e} \left(          -  \log 2          + H(-1;x_2) \right) \nonumber \\
&&         + 2  \log^2 2
          + \frac{\pi^2}{6}
          + G(1/x_2;x_1) H(-1;x_2)
          - G(1/x_2,-1;x_1)
          + G(1/x_2,0;x_1)\nonumber \\
&&
          + 2 G(-1;x_1)  \log 2
          - 2 G(-1;x_1) H(-1;x_2)
          + 2 G(0;x_1)  \log 2
          - 2 G(0;x_1) H(-1;x_2)\nonumber \\
&&
          - 3 G(1, - x_2;x_1)
          - 4 G(1;x_1)  \log 2
          + 3 G(1;x_1) H(-1;x_2)
          - 3 G(1;x_1) H(0;x_2)\nonumber \\
&&
          + 4 G(1,-1;x_1)
          - G(1,0;x_1)
          - 2 H(-1,-1;x_2)
          + 2 H(-1,0;x_2)
          + 2 H(0;x_2)  \log 2\nonumber \\
&&
          - 2 H(0,-1;x_2)
          + 2 H(1;x_2)  \log 2
          - 2 H(1,-1;x_2)
\Bigg] + {\cal O} (\e)\;,
 \\
M_{22} & = & (s_{12})^{-3-2\e} \, S_\Gamma \, \frac{(1-x_1)^{-2\e}(1-x_2)^{-2\e}}{x_1^2 x_2} \Bigg[ \nonumber \\
&& \frac{1}{2\e^2} + \frac{1}{\e} \left(
          \frac{1}{2}G( -  x_2; x_1)
          - \frac{1}{2} G(-1; x_1)
          - G(0; x_1)
          + \frac{1}{2} H(0; x_2)
          \right)
\nonumber \\
&& 
       - \frac{7\pi^2}{24}
          - \frac{1}{2} G( -  x_2;x_1) H(-1; x_2)
          + G( -  x_2; x_1) H(0; x_2)
          + \frac{1}{2} G( -  x_2,-1; x_1)
\nonumber \\
&&           - G( -  x_2,0; x_1)
          - \frac{3}{2} G(-1, -  x_2; x_1)
          - \frac{1}{2} G(-1; x_1)  \log 2
          + G(-1; x_1) H(-1; x_2)
\nonumber \\
&&           - \frac{3}{2} G(-1; x_1) H(0; x_2)
          + G(-1,-1; x_1)
          - G(0, -  x_2; x_1)
          - G(0; x_1) H(0; x_2)
\nonumber \\
&&           + G(0,-1; x_1)
          + 2 G(0,0; x_1)
          - \frac{1}{2} G(1, -  x_2; x_1)
          + \frac{1}{2} G(1; x_1)  \log 2
\nonumber \\
&&          + \frac{1}{2} G(1; x_1) H(-1; x_2)
          - \frac{1}{2} G(1; x_1) H(0; x_2)
          - \frac{1}{2} G(1,-1; x_1)
          - G(1,0; x_1)
\nonumber \\
&&           - 2 H(-1; x_2)  \log 2
          + H(-1,-1; x_2)
          - H(-1,0; x_2)
          - \frac{3}{2} H(0,-1; x_2)
          + H(0,0; x_2)
\nonumber \\
&&           - 2 H(1; x_2)  \log 2
          + 2 H(1,-1; x_2)
          - 3 H(1,0; x_2)
          - g_{22}( x_1, x_2)
 \Bigg] + {\cal O} (\e)\;,
 \\
M_{23} & = & (s_{12})^{-3-2\e} \, S_\Gamma \, \frac{(1-x_1)^{-2\e}(1-x_2)^{-2\e}}{x_1x_2 (1-x_1^2)(x_1+x_2)}  \Bigg[\nonumber \\
&&     
        \frac{1}{\e} \left(
            \frac{1}{2} G( - x_2;x_1)
          - \frac{1}{2} G(-1;x_1)
          + \frac{3}{2} H(0;x_2)
          \right)\nonumber \\
&&     
       - \frac{29\pi^2}{24}
          - G( - x_2, - x_2;x_1)
          - \frac{1}{2} G( - x_2;x_1) H(-1;x_2)
          - 2 G( - x_2;x_1) H(0;x_2)
\nonumber \\
&&               + \frac{3}{2} G( - x_2,-1;x_1)
          - G( - x_2,0;x_1)
          - \frac{5}{2} G(-1, - x_2;x_1)
          + \frac{1}{2} G(-1;x_1)   \log 2
          \nonumber \\
&&     
          - \frac{5}{2} G(-1;x_1) H(0;x_2)
          + 2 G(-1,-1;x_1)
          + G(-1,0;x_1)
          - G(0, - x_2;x_1)
 \nonumber \\
&&              - 3 G(0;x_1) H(0;x_2)
          + G(0,-1;x_1)
          - \frac{3}{2} G(1, - x_2;x_1)
          - \frac{1}{2} G(1;x_1)   \log 2
\nonumber \\
&&               + \frac{1}{2} G(1;x_1) H(-1;x_2)
          + \frac{1}{2} G(1;x_1) H(0;x_2)
          + \frac{3}{2} G(1,-1;x_1)
          - 2 H(-1;x_2)   \log 2
\nonumber \\
&&               - H(-1;x_2) H(1;x_1)
          + H(-1,-1;x_2)
          - 3 H(-1,0;x_2)
          - 2 H(0;x_2)   \log 2
\nonumber \\
&&               + 2 H(0;x_2) H(1;x_1)
          + \frac{5}{2} H(0,-1;x_2)
          - 6 H(0,0;x_2)
          + H(1;x_1)   \log 2
\nonumber \\
&&               - 3 H(1;x_2)   \log 2
          + 3 H(1,-1;x_2)
          - 6 H(1,0;x_2)
          + 4 g_{23b}(x_2)
          - 2 g_{22}(x_1,x_2)
\nonumber \\
&&               + g_{23}(x_1,x_2)
          + g_{25}(x_1,x_2)
        \Bigg] + {\cal O} (\e)\;,
\\
I_{24} & = & (s_{12})^{-3-2\e} \, S_\Gamma \, \frac{(1-x_1)^{-2\e}(1-x_2)^{-2\e}}{x_1x_2 (x_1-x_2)}  \Bigg[   \frac{1}{\e} \big(
          - G(-1; x_1)
          + H(-1; x_2)
          \big)\nonumber \\
           &&
+ \frac{\pi^2}{6}
          + 2 G(1/ x_2; x_1) H(-1; x_2)
          - 2 G(1/ x_2,-1; x_1)
          + 2 G(1/ x_2,0; x_1)
\nonumber \\
&&             - 3 G(-1, -  x_2; x_1)
          + G(-1; x_1)  \log 2
          - 3 G(-1; x_1) H(0; x_2)
          + 2 G(-1,-1; x_1)
\nonumber \\
&&             - 2 G(0; x_1) H(-1; x_2)
          + 2 G(0,-1; x_1)
          - 3 G(1, -  x_2; x_1)
          - G(1; x_1)  \log 2
\nonumber \\
&&             + 3 G(1; x_1) H(-1; x_2)
          - 3 G(1; x_1) H(0; x_2)
          + G(1,-1; x_1)
          - 2 G(1,0; x_1)
\nonumber \\
&&             - 3 H(-1; x_2)  \log 2
          - 2 H(-1,-1; x_2)
          + 3 H(-1,0; x_2)
          - 3 H(1; x_2)  \log 2
\nonumber \\
&&             + 3 H(1,-1; x_2)
          - 3 H(1,0; x_2)
          + 2 g_{24}( x_1; x_2)
\Bigg] + {\cal O} (\e)\;,
\\
M_{25} & = & (s_{12})^{-3-2\e} \, S_\Gamma \, \frac{(1-x_1)^{-2\e}(1-x_2)^{-2\e}}{x_2 (1-x_1^2)}  \Bigg[
\nonumber \\
& & 
         \frac{1}{\e} \left(
           \frac{1}{2} G( - x_2;x_1)
          - \frac{1}{2} G(-1;x_1)
          + \frac{3}{2} H(0;x_2)
          \right)\nonumber \\
&& 
       - \frac{23\pi^2}{24}
          - \frac{1}{2} G( - x_2;x_1) H(-1;x_2)
          + G( - x_2;x_1) H(0;x_2)
          + \frac{1}{2} G( - x_2,-1;x_1)\nonumber \\
&& 
          - G( - x_2,0;x_1)
          - \frac{5}{2} G(-1, - x_2;x_1)
          + \frac{1}{2} G(-1;x_1)   \log 2
          - \frac{5}{2} G(-1;x_1) H(0;x_2)
\nonumber \\
&&           + 2 G(-1,-1;x_1)
          + G(-1,0;x_1)
          + G(0, - x_2;x_1)
          - G(0;x_1) H(0;x_2)
          - G(0,-1;x_1)
\nonumber \\
&&           - \frac{3}{2} G(1, - x_2;x_1)
          - \frac{1}{2} G(1;x_1)   \log 2
          + \frac{1}{2} G(1;x_1) H(-1;x_2)
          + \frac{1}{2} G(1;x_1) H(0;x_2)
\nonumber \\
&&           + \frac{3}{2} G(1,-1;x_1)
          - H(-1;x_2)   \log 2
          - H(-1;x_2) H(1;x_1)
          - H(-1,0;x_2)
\nonumber \\
&&           - 2 H(0;x_2)   \log 2
          + 2 H(0;x_2) H(1;x_1)
          + \frac{5}{2} H(0,-1;x_2)
          - 5 H(0,0;x_2)
          \nonumber \\
&& 
          + H(1;x_1)   \log 2
          - 2 H(1;x_2)   \log 2
          + 2 H(1,-1;x_2)
          - 5 H(1,0;x_2)
\nonumber \\
&&           + 4 g_{23b}(x_2)
          - g_{22}(x_1,x_2)
          + g_{23}(x_1,x_2)
          + g_{25}(x_1,x_2)
\Bigg] +  {\cal O} (\e)\;,
\\
I_{27} & = & (s_{12})^{-3-2\e} \, S_\Gamma \, \frac{(1-x_1)^{-2\e}(1-x_2)^{-2\e}}{x_2 (1-x_1^2)}  \Bigg[\nonumber  \\
& & -\frac{1}{\e^2} + \frac{1}{\e} \big(
          - 2 G( - x_2;x_1)
          + 2 G(-1;x_1)
          - 2 H(0;x_2)
          \big)
\nonumber \\
&&     
       + \frac{3\pi^2}{2}
          - 2G(1/x_2;x_1)H(-1;x_2)
          + 2G(1/x_2,-1;x_1)
          - 2G(1/x_2,0;x_1)
\nonumber \\
&&               + 4G(-1, - x_2;x_1)
          + 4G(-1;x_1)H(0;x_2)
          - 4G(-1,-1;x_1)
          + 2G(0;x_1)H(-1;x_2)
\nonumber \\
&&               - 2G(0,-1;x_1)
          + 2G(0,0;x_1)
          + 2H(-1;x_2) \log 2
          + 2H(0,0;x_2)
\nonumber \\
&&               + 2H(1;x_2) \log 2
          - 2H(1,-1;x_2)
          + 6H(1,0;x_2)
\Bigg]  + {\cal O} (\e)\;,
\\
I_{28} & = & (s_{12})^{-3-2\e} \, S_\Gamma \, \frac{(1-x_1)^{-2\e}(1-x_2)^{-2\e}}{x_1 x_2 (1-x_1^2)}  \Bigg[ \nonumber  \\
&& -\frac{1}{2\e^2} + \frac{1}{\e} \left(
           2   \log 2
          - G(-1;x_1)
          + G(0;x_1)
          - H(-1;x_2)
          + H(0;x_2) \right) \nonumber \\
&& 
          - 4   \log^2 2
          - \frac{2\pi^2}{3}
          - 3  G(-1, - x_2;x_1)
          + G(-1;x_1)  H(-1;x_2)
          - G(-1;x_1)  H(0;x_2) \nonumber \\
&& 
          + 2  G(-1,-1;x_1)
          + G(-1,0;x_1)
          - 4  G(0;x_1)   \log 2
          + 2  G(0;x_1)  H(-1;x_2) \nonumber \\
&& 
          - 2  G(0;x_1)  H(0;x_2)
          + 2  G(0,-1;x_1)
          - 2  G(0,0;x_1)
          + 3  G(1, - x_2;x_1)\nonumber \\
&& 
          + 4  G(1;x_1)   \log 2         
            - 3  G(1;x_1)  H(-1;x_2)
          + 3  G(1;x_1)  H(0;x_2)
          - 4  G(1,-1;x_1) \nonumber \\
&& 
          + G(1,0;x_1)
                    + 2  H(-1,-1;x_2)
          - 2  H(-1,0;x_2)
          - 4  H(0;x_2)   \log 2
          + 2  H(0,-1;x_2)\nonumber \\
&& 
          - 2  H(0,0;x_2) 
                    - 4  H(1;x_2)   \log 2
          + 4  H(1,-1;x_2)
          - 4  H(1,0;x_2)
          \Bigg] \nonumber \\
&& + (s_{12})^{-3-2\e} \, S_\Gamma \, \frac{(1-x_1)^{-2\e}(1-x_2)^{-2\e}}{x_2 (1-x_1^2)}  \Bigg[ \nonumber  \\
&& - \frac{3}{2\e^2} + \frac{1}{\e} \left( 
          - 2  G( - x_2;x_1)
          + G(-1;x_1)
          + G(0;x_1)
          + H(-1;x_2)
          - H(0;x_2)
\right) \nonumber \\
&&
       + \pi^2
          + G(-1, - x_2;x_1)
          + 4  G(-1;x_1)   \log 2
          - 3  G(-1;x_1)  H(-1;x_2)\nonumber \\
&&
          + 3  G(-1;x_1)  H(0;x_2)
          - 2  G(-1,-1;x_1)
          + G(-1,0;x_1)
          + 2  G(0, - x_2;x_1)\nonumber \\
&&
          - 2  G(0,-1;x_1)
          - 3  G(1, - x_2;x_1)
          - 4  G(1;x_1)   \log 2
          + 3  G(1;x_1)  H(-1;x_2)\nonumber \\
&&
          - 3  G(1;x_1)  H(0;x_2) 
           + 4  G(1,-1;x_1)
          - G(1,0;x_1)
          + 2  H(-1;x_2)   \log 2\nonumber \\
&&
          - 2  H(-1,-1;x_2)
                    + 2  H(-1,0;x_2)
          + 2  H(1;x_2)   \log 2
          - 2  H(1,-1;x_2)\nonumber \\
&&
          + 2  H(1,0;x_2)
           \Bigg] + {\cal O} (\e)\;,
\\
I_{29} & = & (s_{12})^{-3-2\e} \, S_\Gamma \, \frac{(1-x_1)^{-2\e}(1-x_2)^{-2\e}}{x_1x_2 (1+x_1) (1+x_2)}  \Bigg[ 
\nonumber \\
& & \frac{1}{2\e^2} + \frac{1}{\e} \left(
          - 2  \log 2
          + G(-1;x_1)
          - G(0;x_1)
          + H(-1;x_2)
          - H(0;x_2)
\right) \nonumber \\
 &&         + 4  \log^2 2
          + \frac{2\pi^2}{3}
          + 3 G(-1, - x_2;x_1)
          - G(-1;x_1) H(-1;x_2)
          + G(-1;x_1) H(0;x_2)\nonumber \\
&& 
          - 2 G(-1,-1;x_1)
          - G(-1,0;x_1)
          + 4 G(0;x_1)  \log 2
          - 2 G(0;x_1) H(-1;x_2)\nonumber \\
&& 
          + 2 G(0;x_1) H(0;x_2)
          - 2 G(0,-1;x_1)
          + 2 G(0,0;x_1)
          - 3 G(1, - x_2;x_1)\nonumber \\
&& 
          - 4 G(1;x_1)  \log 2
          + 3 G(1;x_1) H(-1;x_2)
          - 3 G(1;x_1) H(0;x_2)
          + 4 G(1,-1;x_1)\nonumber \\
&& 
          - G(1,0;x_1)
          - 2 H(-1,-1;x_2)
          + 2 H(-1,0;x_2)
          + 4 H(0;x_2)  \log 2
          - 2 H(0,-1;x_2)\nonumber \\
&& 
          + 2 H(0,0;x_2)
          + 4 H(1;x_2)  \log 2
          - 4 H(1,-1;x_2)
          + 4 H(1,0;x_2)
\Bigg]\nonumber \\
&& + (s_{12})^{-3-2\e} \, S_\Gamma \, \frac{(x_1+x_2)(1-x_1)^{-2\e}(1-x_2)^{-2\e}}{x_1x_2 (1-x_1^2) (1-x_2^2)} \Bigg[ 
\nonumber \\
&& \frac{3}{\e^2} 
+\frac{2}{\e} \left(
          - \log 2
          + 3 G( -  x_2 ;x_1)
          -  G(-1 ;x_1)
          - 2 G(0 ;x_1)
          -  H(-1 ;x_2)
          +  H(0 ;x_2)
          \right)\nonumber \\ &&
       + 4 \log^2 2
          - \frac{4\pi^2}{3}
          + \log 2 \left(
          - 4 G(-1 ;x_1)
          + 4 G(0 ;x_1)
          - 4 H(-1 ;x_2) 
          + 4 H(0 ;x_2) \right)\nonumber \\ &&
          + 12 G( -  x_2, -  x_2 ;x_1)
          - 6 G( -  x_2 ;x_1) H(-1 ;x_2)
          + 6 G( -  x_2 ;x_1) H(0 ;x_2)\nonumber \\ &&
          - 6 G( -  x_2,-1 ;x_1)
          - 6 G( -  x_2,0 ;x_1)
          - 6 G(-1, -  x_2 ;x_1)
          + 6 G(-1 ;x_1) H(-1 ;x_2)\nonumber \\ &&
          - 6 G(-1 ;x_1) H(0 ;x_2)
          + 4 G(-1,-1 ;x_1)
          + 2 G(-1,0 ;x_1)
          - 6 G(0, -  x_2 ;x_1)\nonumber \\ &&
          + 2 G(0,-1 ;x_1)
          + 4 G(0,0 ;x_1)
          + 4 H(-1,-1 ;x_2)
          - 4 H(-1,0 ;x_2)
          - 4 H(0,-1 ;x_2)\nonumber \\ &&
          + 4 H(0,0 ;x_2)
\Bigg]+ {\cal O} (\e)\;.
\end{eqnarray}
In the above equations, we introduced the functions:
\begin{eqnarray}
z (x_1,x_2)&=& \sqrt{1 + 4x_1x_2 + 4 x_2^2} \;, \\[1mm]
g_{13} (x_1,x_2) & = & \frac{1}{z(x_1,x_2)} \left[ \log\left(\frac{ z(x_1,x_2)+ 1 - 2 x_2}
{z(x_1,x_2) - 1 + 2 x_2}\right) + 
 \log\left(\frac{z(x_1,x_2)+1}{z(x_1,x_2)-1}\right)\right] \;, \\[1mm]
g_{22} (x_1,x_2) & = & \int_1^{x_1} \d y_1 \,\frac{g_{13}(y_1,x_2)}{y_1+x_2}\\ 
&=& 
\frac{\pi ^2}{12}+\frac{3}{2}{  \log}^2 x_2+{  \log}2\, {  \log}\left(\frac{1-{x_2}}{1+{x_2}}\right)-2 {  \log}{x_2} \,{  \log}(1+{x_2})+\frac{1}{2} {  \log}(1+{x_2})^2\nonumber \\ &&
+{  \log}\left(\frac{1-{x_2}}{{x_2}}\right) {  \log}(z-1)-\frac{1}{2} {  \log}^2\left(\frac{z-1}{z+1}\right)+{  \log}({x_2}) {  \log}(1-2 {x_2}+z)\nonumber \\ &&
-{  \log}(1-{x_2}) {  \log}(z-1+2 {x_2})+{  \log}\left(\frac{z+1}{2}\right) {  \log}\left(\frac{z-1+2 {x_2}}{z+1-2 {x_2}}\right)\nonumber \\ &&
+{\Li}_2\left(-\frac{1}{{x_2}}\right)-{\Li}_2\left(-\frac{4 {x_2}}{2-2 {x_2}}\right)+{\Li}_2\left(-\frac{2 {x_2}}{2-2 {x_2}}\right)\nonumber \\ &&
-{\Li}_2\left(\frac{1-z}{2-2 {x_2}}\right)+{\Li}_2\left(\frac{1-z}{2 {x_2}}\right)+{\Li}_2\left(\frac{1-2 {x_2}-z}{2-2 {x_2}}\right)\nonumber\\ &&-{\Li}_2\left(-\frac{1-2 {x_2}+z}{2 {x_2}}\right)\;,
\\[1mm]
g_{23} (x_1,x_2) &=& \int_1^{x_1} \d y_1\, \Big(g_{13}(y_1,x_2)-g_{13}(1,x_2)\Big)
\, \frac{-1-2x_2}{1-y_1}\\
&=& 
 \log 2\, \log\left(1 + \frac{1}{2 x_2}\right) + \log\left(2 + \frac{1}{x_2}\right) \log(2 (1 + x_2)) 
 \nonumber\\ && + 
 \log(2 x_2) \left(\log(1 + x_2)- \log(1 + 2 x_2)\right) - \log(4 x_2) \log(1 + 2 x_2) \nonumber\\ && + 
 \log(-1 + 2 x_2 + z) \log\left(\frac{1 + 2 x_2 + z}{2}\right) \nonumber\\ && + 
 \log(1 + z) \left(\log 2 + \log(x_2) - \log(1 + 2 x_2 + z)\right) \nonumber\\ && + 
 \log(1 - 2 x_2 + z) \left(\log 4 + \log(x_2) - 
    \log(1 + 2 x_2 + z)\right) \nonumber\\ && + 
 \log\left(\frac{1+x_2}{2(z-1)x_2^2}\right)
 \log\left(\frac{2 + 4 x_2}{1 + 2 x_2 + z}\right) \nonumber\\ && +  
 \Li_2\left( -\frac{1}{2 x_2}\right) - \Li_2( -2 x_2) + 
 \Li_2\left( -\frac{1 + x_2}{x_2}\right)- \Li_2\left( -\frac{x_2}{1+x_2}\right)  \nonumber\\ &&+ 
 \Li_2\left( \frac{1 - z}{2 + 2 x_2}\right) + 
 \Li_2\left( \frac{1 - 2 x_2 - z}{2}\right) + 
 \Li_2\left( \frac{1 + 2 x_2 - z}{4 x_2}\right)   \nonumber\\ &&+ 
 \Li_2\left( \frac{1 + 2 x_2 - z}{2 x_2}\right) - 
 \Li_2\left( \frac{1 + 2 x_2 - z}{2 + 2 x_2}\right)   - 
 \Li_2\left( \frac{1 + 2x_2 -  z}{2}\right)\nonumber\\ && - 
 \Li_2\left( -\frac{1 + z}{2 x_2}\right) - 
 \Li_2\left( -\frac{1 - 2 x_2 + z}{4 x_2}\right)
\;,
 \\[1mm]
g_{23b} (x_2) &=& \int_1^{x_2} \d y_2 \frac{1}{1+2y_2} \, \Big( H(-1;y_2)  - \log 2 -2 H(0;y_2) \Big)\\
&=& \frac{\pi ^2}{6}+ \log(2x_2) \, \log(1+2x_2)+\frac{1}{2} \Li_2(-1-2 x_2)+ \Li_2(-2 x_2)\;,
 \\[1mm]
 g_{23c} (x_2) &=& \int_1^{x_2} \d y_2 \left(\frac{1}{y_2} -\frac{1}{1+y_2} \right) g_{23b} (y_2) \\
 & = & 
 -\frac{\pi^2}{12}\ \log 2 + \frac{17}{6} \log^3 2 + \frac{\pi^2}{6} \log 3 - 
 2 \log^2 2 \log 3 + \frac{1}{6} \log^3 3 + \frac{\pi^2}{6} \log(x_2)\nonumber \\ && - 
 \frac{1}{4} \log^2 2 \log(x_2) - \frac{1}{12} \log^3(x_2) + \frac{\pi^2}{12} \log(1 + x_2) + 
 \frac{7}{4} \log^2 2 \log(1 + x_2)\nonumber \\ && - \frac{1}{2} \log 2 \log(x_2) \log(1 + x_2) - 
 \frac{1}{4} \log^2(x_2) \log(1 + x_2) + 2 \log 2 \log^2(1 + x_2)\nonumber \\ && + 
 \frac{7}{4} \log(x_2) \log^2(1 + x_2) - \frac{1}{12} \log^3(1 + x_2) + 
 \frac{\pi^2}{6} \log(1 + 2 x_2) \nonumber \\ &&- \frac{1}{2} \log^2 2 \log(1 + 2 x_2) + 
 \frac{1}{2} \log 2 \log(x_2) \log(1 + 2 x_2)\nonumber \\ && + \frac{1}{2} \log^2(x_2) \log(1 + 2 x_2) - 
 \frac{3}{2} \log 2 \log(1 + x_2) \log(1 + 2 x_2) \nonumber \\ &&- 
 2 \log(x_2) \log(1 + x_2) \log(1 + 2 x_2) - \log^2(1 + x_2) \log(1 + 2 x_2)\nonumber \\ && + 
 \frac{3}{4} \log 2 \log^2(1 + 2 x_2) + \log(x_2) \log^2(1 + 2 x_2) + 
 \frac{7}{4} \log(1 + x_2) \log^2(1 + 2 x_2)\nonumber \\ && - \log^3(1 + 2 x_2) + 
 \frac{1}{2} \log 2 \log\left(\frac{1 + 2 x_2}{2 + 2 x_2}\right) \log(2 + 2 x_2) \nonumber \\ &&+ 
 \frac{1}{2} \log(1 + x_2) \log\left(\frac{1 + 2 x_2}{2 + 2 x_2}\right) \log(2 + 2 x_2) + 
 \frac{3}{2} \log(x_2) \Li_2\left( \frac{x_2}{1 + x_2}\right) \nonumber \\ &&- 
 \frac{3}{2} \log(1 + x_2) \Li_2\left( \frac{x_2}{1 + x_2}\right) + 
 \log(x_2) \Li_2\left( \frac{x_2}{1 + 2 x_2}\right) \nonumber \\ &&- 
 \log(1 + 2 x_2) \Li_2\left( \frac{x_2}{1 + 2 x_2}\right) + 
 \frac{1}{2} \log 2 \,\Li_2\left( \frac{2 x_2}{1 + 2 x_2}\right) \nonumber \\ &&- 
 \frac{1}{2} \log(x_2) \Li_2\left( \frac{2 x_2}{1 + 2 x_2}\right) + 
 \log(1 + x_2) \Li_2\left( \frac{2 x_2}{1 + 2 x_2}\right)\nonumber \\ && - 
 \frac{1}{2} \log(x_2) \Li_2\left( \frac{1 + x_2}{1 + 2 x_2}\right) + 
 \frac{3}{2} \log(1 + x_2) \Li_2\left( \frac{1 + x_2}{1 + 2 x_2}\right) \nonumber \\ &&- 
 \log(1 + 2 x_2) \Li_2\left( \frac{1 + x_2}{1 + 2 x_2}\right) + 
 \frac{1}{2} \log 2 \, \Li_2\left( \frac{1 + 2 x_2}{2 + 2 x_2}\right) \nonumber \\ &&- 
 \frac{1}{2} \log(x_2) \Li_2\left( \frac{1 + 2 x_2}{2 + 2 x_2}\right) + 
 \log(1 + x_2) \Li_2\left(\frac{ 1 + 2 x_2}{2 + 2 x_2}\right) \nonumber \\ &&- 
 2 \Li_3\left( -\frac{1}{2}\right) - \Li_3\left( -\frac{1}{3}\right) - \frac{1}{2} \Li_3\left( \frac{1}{4}\right) - 
 \Li_3\left( \frac{3}{4}\right) + \Li_3\left( -1 - 2 x_2\right)\nonumber \\ &&
  + \frac{3}{2} \Li_3\left( -2 x_2\right) + 
 \Li_3\left( -x_2\right) + \Li_3\left( \frac{1}{1 + x_2}\right) - 
 \frac{1}{2} \Li_3\left( \frac{x_2}{1 + x_2}\right) \nonumber \\ &&
 + \frac{1}{2} \Li_3\left( -\frac{1 + x_2}{x_2}\right) + 
 \Li_3\left( \frac{1}{1 + 2 x_2}\right) - \Li_3\left( \frac{x_2}{1 + 2 x_2}\right) + 
 \Li_3\left( \frac{2 x_2}{1 + 2 x_2}\right) \nonumber \\ &&- \Li_3\left( \frac{1 + x_2}{1 + 2 x_2}\right) + 
 \frac{1}{2} \Li_3\left( \frac{1}{2 + 2 x_2}\right) 
 + \Li_3\left( \frac{1 + 2 x_2}{2 + 2 x_2}\right) + \frac{5}{16}\zeta_3\;,
  \\[1mm]
g_{24} (x_1,x_2) &=&\int_1^{x_1} \d y_1 \, \frac{\log(1+y_1)-\log(1+x_2)}{y_1-x_2} \\
&=& 
\frac{ { \ \log^2}2}{2}-\frac{1}{2}  { \ \log^2}(1+ { x_1})- { \ \log}2\,  { \ \log}(1- { x_2})+ { \ \log}(1+ { x_1})  { \ \log}( { x_1}- { x_2})\nonumber \\ &&
+ { \ \log}(1- { x_2}) \,{ \ \log}(1+ { x_2})- { \ \log}( { x_1}- { x_2}) \,{ \ \log}(1+ { x_2})+ {\Li_2}\left(\frac{1+ { x_2}}{2}\right)\nonumber \\ &&- {\Li_2}\left(\frac{1+ { x_2}}{1+ { x_1}}\right)\;,
 \\[1mm]
g_{25} (x_1,x_2)  &=& \int_1^{x_1} \d y_1\, g_{13}(y_1,x_2)
\, \frac{1-2x_2}{1+y_1}\\ 
&=&-\frac{\pi^2}{12} + \frac{\log^2 2}{2} + \log\left(-1 + \frac{1}{x_2}\right) \log(2 - 2 x_2) - 
 \log^2(2 - 2 x_2) \nonumber \\ &&
 + \log(2 - 2 x_2) \log(x_2) + \frac{1}{2} \log(x_2) \log(4 x_2) - 
 \log(2 - 2 x_2) \log( z-1) \nonumber \\ && + \log(2 x_2) \log( z-1) - 
 \log(2 x_2) \log\left(\frac{1 - 2 x_2 + z}{2}\right) \nonumber \\ &&+ 
 \log(z-1) \log(1 - 2 x_2 + z) - 
 \frac{1}{2}\log^2(1 - 2 x_2 + z)\nonumber \\ &&+ 
 \log(2 - 2 x_2) \log(-1 + 2 x_2 + z) - 
 \log( z-1) \log(-1 + 2 x_2 + z) \nonumber \\ &&+ 
 \log(1 - 2 x_2 + z) \log(-1 + 2 x_2 + z) - 
 \frac{1}{2} \log^2(-1 + 2 x_2 + z)\nonumber \\ && - \Li_2\left( -\frac{1}{x_2}\right) - 
 \Li_2\left( -\frac{x_2}{1 - x_2}\right) + \Li_2\left(-\frac{ 2 x_2}{1 - x_2}\right) + 
 \Li_2\left( -\frac{ z-1}{2 (1 - x_2)}\right) \nonumber \\ &&- 
 \Li_2\left( -\frac{z-1}{2 x_2}\right) + 
 \Li_2\left( -\frac{- 2 x_2 + z+1}{2 x_2}\right) - 
 \Li_2\left( -\frac{ 2 x_2 + z-1}{2 (1 - x_2)}\right)\,,
\end{eqnarray}
with $z \equiv z(x_1,x_2)$.

\subsection{Collinear $x_1$ region}
In the collinear $x_1$ region, we can expand around $x_2=1$, and the master integrals read:
\begin{eqnarray} 
I_{13} &=& {\cal O} \big((1-x_2)\big)\;, \\
I_{16} & = & 
(s_{12})^{-3-2\e} \, S_\Gamma \, \frac{(1-x_1)^{-2\e}(1-x_2)^{-1-2\e}}
{2 x_1} \Bigg[ \nonumber \\
 &&      -\frac{1}{\e^2}
       + \frac{1}{\e} (
          2 \log 2
          - 2 H(-1;x_1)
          ) 
 + \bigg(
       - 2 \log^2 2
          + \frac{7\pi^2}{6}
          + 2 H(-1;x_1) \log 2\nonumber \\ &&
          - 2 H(-1,-1;x_1)
          + 4 H(0,0;x_1)
          - 2 H(1;x_1) \log 2
          + 2 H(1,-1;x_1)
          + 4 H(1,0;x_1)
           \bigg)\nonumber \\ &&
       + \e   \bigg(
           \frac{33}{4} \zeta_3
          + \frac{4}{3} \log^3 2
          +  \log^2 2 \left( - H(-1;x_1)
          + 3 H(1;x_1) \right)
           + \log 2 \big( 2 H(-1,-1;x_1)\nonumber \\ && 
          + 4 H(-1,0;x_1) 
          - 8 H(0,0;x_1) 
          - 2 H(1,-1;x_1) 
          - 4 H(1,0;x_1) 
          + 4 H(1,1;x_1) \big)\nonumber \\ &&
         + \frac{7}{3} \pi^2 (-\log 2
          - H(1;x_1) 
          +  H(-1;x_1) 
          -  H(0;x_1) )
          - 2 H(-1,-1,-1;x_1)\nonumber \\ &&
          - 4 H(-1,0,-1;x_1)
          + 8 H(-1,0,0;x_1)
          + 8 H(-1,1,0;x_1)
          + 4 H(0,-1,0;x_1)\nonumber \\ &&
          + 8 H(0,0,-1;x_1)
          - 16 H(0,0,0;x_1)
          - 12 H(0,1,0;x_1)
           + 2 H(1,-1,-1;x_1)\nonumber \\ &&
          + 4 H(1,-1,0;x_1)
          + 4 H(1,0,-1;x_1)
          - 16 H(1,0,0;x_1)\nonumber \\ &&
          - 4 H(1,1,-1;x_1)
          - 12 H(1,1,0;x_1)
          \bigg)
          + {\cal O} \big(  (1-x_2) \big)
\Bigg] + {\cal O} (\e^2)\;,\\
I_{17} & = & 
(s_{12})^{-3-2\e} \, S_\Gamma \, \frac{(1-x_1)^{-2\e}(1-x_2)^{-2\e}}
{x_1(1-x_1)} \Bigg[ \nonumber \\
&&     -\frac{1}{2\e^2} + \frac{1}{\e}  \left(     
           \log 2
          - H(-1;x_1)
          + H(0;x_1)
          \right)
       +  \bigg(
       -\frac{\pi^2}{6}
          - 2 \log^2 2
          + \log 2\big(
           H(-1;x_1)  \nonumber \\ &&
          - 2 H(0;x_1) 
          - H(1;x_1) \big)
          - H(-1,-1;x_1)
          + H(-1,0;x_1)\nonumber \\ &&
          + 2 H(0,-1;x_1)
          - 2 H(0,0;x_1)
          + H(1,-1;x_1)
           - H(1,0;x_1)
          \bigg)\nonumber \\ &&
       + \e  \bigg(
            \zeta_3
          + \frac{2}{3} \log^3 2
          + \frac{1}{2} \log^2 2 \left (
          -  H(-1;x_1) 
          + 4 H(0;x_1) 
          + 3 H(1;x_1) \right) \nonumber \\ &&
          + \log 2\Big( 
           H(-1,-1;x_1) 
          -  H(-1,0;x_1) 
          - 2 H(0,-1;x_1) 
          + 4 H(0,0;x_1) 
          + 2 H(0,1;x_1) \nonumber \\ &&
          - H(1,-1;x_1) 
          + 3 H(1,0;x_1) 
          + 2 H(1,1;x_1) \Big)
          + \frac{\pi^2}{3} \left(
            H(-1;x_1)
          + H(1;x_1)\right)\nonumber \\ &&
          - H(-1,-1,-1;x_1)
          + H(-1,-1,0;x_1)
          + H(-1,0,-1;x_1)
          - H(-1,0,0;x_1)\nonumber \\ &&
          + 2 H(0,-1,-1;x_1)
          - 2 H(0,-1,0;x_1)
          - 4 H(0,0,-1;x_1)
          + 4 H(0,0,0;x_1)\nonumber \\ &&
          - 2 H(0,1,-1;x_1)
          + 2 H(0,1,0;x_1)
          + H(1,-1,-1;x_1)
          - H(1,-1,0;x_1)\nonumber \\ &&
          - 3 H(1,0,-1;x_1)
          + 3 H(1,0,0;x_1)
          - 2 H(1,1,-1;x_1)
          + 2 H(1,1,0;x_1)
          \bigg)\nonumber \\ &&
         + {\cal O} \big(  (1-x_2) \big)
\Bigg]  + {\cal O} (\e^2)\;,\\
I_{22} & = & 
(s_{12})^{-3-2\e} \, S_\Gamma \, \frac{(1-x_1)^{-2\e}(1-x_2)^{-2\e}}
{x_1^2} \Bigg[ \nonumber \\
&&       \frac{1}{2\e^2}        - \frac{1}{\e}  H(0;x_1)
          + \bigg(
       - \frac{1}{2} \log^2 2
          - H(-1,0;x_1)
          + 2 H(0,0;x_1) \nonumber \\ &&
          - H(1;x_1) \log 2
          + H(1,-1;x_1)
          + H(1,0;x_1)\bigg)\nonumber \\ &&
       + \e  \bigg(
          - \zeta_3
          + \frac{1}{2} \log^3 2
          + \frac{1}{2} \log^2 2 \,\left(2H(0;x_1) +  3 H(1;x_1) \right)\nonumber \\ &&
          +  \log 2 \left( H(-1,0;x_1)
          + 2 H(0,1;x_1)
          - H(1,-1;x_1) 
          + H(1,0;x_1)
          + 2 H(1,1;x_1) \right)\nonumber \\ &&
          + \frac{\pi^2}{3} \left( H(-1;x_1) - H(1;x_1) \right)
          - H(-1,-1,0;x_1)
          - H(-1,0,-1;x_1)\nonumber \\ &&
          + 3 H(-1,0,0;x_1)
          + 2 H(-1,1,0;x_1)
          + 2 H(0,-1,0;x_1)
          - 4 H(0,0,0;x_1)\nonumber \\ &&
          - 2 H(0,1,-1;x_1)
          - 2 H(0,1,0;x_1)
          + H(1,-1,-1;x_1)
          + H(1,-1,0;x_1) \nonumber \\ &&
          - H(1,0,-1;x_1)
          - 3 H(1,0,0;x_1)
          - 2 H(1,1,-1;x_1)\nonumber \\ &&
          - 2 H(1,1,0;x_1)
         \bigg ) + {\cal O} \big(  (1-x_2) \big)
\Bigg] + {\cal O} (\e^2)\;,\\
I_{23} &=& {\cal O} \big((1-x_2)\big)\;, \\
I_{24} & = & -
(s_{12})^{-3-2\e} \, S_\Gamma \, \frac{(1-x_1)^{-1-2\e}(1-x_2)^{-2\e}}
{x_1} \Bigg[ \nonumber \\
 &&       \frac{1}{\e}   (
          \log 2
          - H(-1;x_1) )
+\bigg(
       - \frac{3}{2} \log^2 2
          +\log 2 \left(H(-1;x_1) 
          - 2 H(0;x_1) 
          - 2 H(1;x_1)\right)  \nonumber \\ &&
          + 2 H(0,-1;x_1)
          - H(-1,-1;x_1)
          + 2 H(1,-1;x_1)
     \bigg)\nonumber \\ &&
       + \e   \bigg(
           \frac{7}{6} \log^3 2
          + \frac{1}{2} \log^2 2 \left( -H(-1;x_1) 
          + 6 H(0;x_1) 
          + 6 H(1;x_1) \right)
          + \log 2\big( 
          H(-1,-1;x_1) \nonumber \\ &&
          - 2 H(0,-1;x_1) 
          + 4 H(0,0;x_1) 
          + 4 H(0,1;x_1) 
          - 2 H(1,-1;x_1) 
          + 4 H(1,0;x_1) \nonumber \\ &&
          + 4 H(1,1;x_1) \big)
          + \frac{2\pi^2}{3} H(-1;x_1) 
          - H(-1,-1,-1;x_1)
          + 2 H(-1,0,0;x_1)\nonumber \\ &&
          + 2 H(-1,1,0;x_1)
          + 2 H(0,-1,-1;x_1)
          - 4 H(0,0,-1;x_1)
          - 4 H(0,1,-1;x_1)\nonumber \\ &&
          + 2 H(1,-1,-1;x_1)
          - 4 H(1,0,-1;x_1)
          - 4 H(1,1,-1;x_1)
          \bigg)\nonumber \\&&
          + {\cal O} \big(  (1-x_2) \big)
\Bigg] + {\cal O} (\e^2) \;,\\
I_{25} &=& {\cal O} \big((1-x_2)\big) \;,\\
I_{27} & = & 
(s_{12})^{-3-2\e} \, S_\Gamma \, \frac{(1-x_1)^{-2\e}(1-x_2)^{-2\e}}
{(1-x_1^2)} \Bigg[ \nonumber \\
&&       -\frac{1}{\e^2} 
       +\bigg(
         \log^2 2
           + 2 \log 2 \left(H(0;x_1)+  H(1;x_1)\right)
          + \frac{2\pi^2}{3}
          - 2 H(0,-1;x_1)
          + 2 H(0,0;x_1)\nonumber \\ &&
          - 2 H(1,-1;x_1)
          + 2 H(1,0;x_1)
         \bigg)
       + \e   \bigg(
            6 \zeta_3
          - \log^3 2
          - 3 \log^2 2 \left( H(0;x_1) + H(1;x_1) \right)\nonumber \\ &&
         + \log 2 \big(
          2 H(0,-1;x_1)
          - 6 H(0,0;x_1) 
          - 4 H(0,1;x_1) 
          + 2 H(1,-1;x_1) \nonumber \\ &&
          - 6 H(1,0;x_1) 
          - 4 H(1,1;x_1) \big) \log 2
          - \frac{2\pi^2}{3} \left(\log 2
          + H(0;x_1) 
           + H(1;x_1) \right)\nonumber \\ &&
          - 2 H(0,-1,-1;x_1)
          + 2 H(0,-1,0;x_1)
          + 6 H(0,0,-1;x_1)
          - 6 H(0,0,0;x_1)\nonumber \\ &&
          + 4 H(0,1,-1;x_1)
          - 4 H(0,1,0;x_1)
          - 2 H(1,-1,-1;x_1)
          + 2 H(1,-1,0;x_1)\nonumber \\ &&
          + 6 H(1,0,-1;x_1)
          - 6 H(1,0,0;x_1)
          + 4 H(1,1,-1;x_1)
          - 4 H(1,1,0;x_1)
          \bigg)\nonumber \\ &&
          + {\cal O} \big(  (1-x_2) \big)
\Bigg] + {\cal O} (\e^2)\;,\\
I_{28} & = & 
(s_{12})^{-3-2\e} \, S_\Gamma \, \frac{(1-x_1)^{-2\e}(1-x_2)^{-2\e}}
{x_1(1-x_1)} \Bigg[ \nonumber \\
&&  -\frac{1}{2\e^2} + \frac{1}{\e}  \left(     
           \log 2
          - H(-1;x_1)
          + H(0;x_1)
          \right)
       +  \bigg(
          - 2 \log^2 2
          + \log 2\big(
           H(-1;x_1)  \nonumber \\ &&
          - 2 H(0;x_1) 
          - H(1;x_1) \big)
          - H(-1,-1;x_1)
          + H(-1,0;x_1)\nonumber \\ &&
          + 2 H(0,-1;x_1)
          - 2 H(0,0;x_1)
          + H(1,-1;x_1)
           - H(1,0;x_1)
          \bigg)\nonumber \\ &&
       + \e  \bigg(
             \zeta_3
          + \frac{2}{3} \log^3 2
          + \frac{1}{2} \log^2 2 \left (
          -  H(-1;x_1) 
          + 4 H(0;x_1) 
          + 3 H(1;x_1) \right) \nonumber \\ &&
          + \log 2\Big(
            H(-1,-1;x_1) 
          - H(-1,0;x_1)  
          - 2 H(0,-1;x_1) 
           \nonumber \\ &&
          + 4 H(0,0;x_1) 
          + 2 H(0,1;x_1)  
          - H(1,-1;x_1)
                    + 3 H(1,0;x_1) 
          + 2 H(1,1;x_1) \Big) \nonumber \\ &&
          + \frac{\pi^2}{3} \left(
             H(-1;x_1)
          + H(1;x_1)\right)
          -   H(-1,-1,-1;x_1)
          +  H(-1,-1,0;x_1) \nonumber \\ &&     
          + H(-1,0,-1;x_1)
          -  H(-1,0,0;x_1)
          + 2 H(0,-1,-1;x_1)  
          - 2 H(0,-1,0;x_1) \nonumber \\ &&     
          - 4 H(0,0,-1;x_1)
          + 4 H(0,0,0;x_1)
          - 2 H(0,1,-1;x_1)
          + 2 H(0,1,0;x_1)\nonumber \\ &&
          + H(1,-1,-1;x_1)
          - H(1,-1,0;x_1)
          - 3 H(1,0,-1;x_1)
          + 3 H(1,0,0;x_1)\nonumber \\ &&
          - 2 H(1,1,-1;x_1)
          + 2 H(1,1,0;x_1)
          \bigg)
         + {\cal O} \big(  (1-x_2) \big)
\Bigg]\nonumber \\
&&+ (s_{12})^{-3-2\e} \, S_\Gamma \, \frac{(1-x_1)^{-2\e}(1-x_2)^{-2\e}}
{1-x_1^2} \Bigg[ \nonumber \\
&&-\frac{1}{\e^2}
+\bigg(        \log^2 2
          + \frac{2\pi^2}{3}
          + \log 2\left(2 H(0;x_1) 
          + 2 H(1;x_1) \right)
          - 2 H(0,-1;x_1)\nonumber \\ &&
          + 2 H(0,0;x_1)
          - 2 H(1,-1;x_1)
          + 2 H(1,0;x_1)\bigg)\nonumber \\ &&
       + \e   \bigg(
            6 \zeta_3
          - \log^3 2
          - 3 \log^2 2 (H(0;x_1) +H(1;x_1) )
          + \log 2 \big(2 H(0,-1;x_1) \nonumber \\ &&
          - 6 H(0,0;x_1) 
          - 4 H(0,1;x_1) 
          + 2 H(1,-1;x_1) 
          - 6 H(1,0;x_1) 
          - 4 H(1,1;x_1) \big)   \nonumber \\ &&      
	+ \frac{\pi^2}{3} \left(
          -  2 \log 2 
          -  2 H(0;x_1) 
          - 2 H(1;x_1) \right)
           - 2 H(0,-1,-1;x_1)
          + 2 H(0,-1,0;x_1) \nonumber \\ &&    
          + 6 H(0,0,-1;x_1)
          - 6 H(0,0,0;x_1)
          + 4 H(0,1,-1;x_1)
          - 4 H(0,1,0;x_1) \nonumber \\ &&    
          - 2 H(1,-1,-1;x_1)
          + 2 H(1,-1,0;x_1)
          + 6 H(1,0,-1;x_1)
          - 6 H(1,0,0;x_1) \nonumber \\ &&    
          + 4 H(1,1,-1;x_1)
          - 4 H(1,1,0;x_1)
            \bigg)     + {\cal O} \big(  (1-x_2) \big)
\Bigg] + {\cal O} (\e^2)\;,\\
I_{29} & = & 
(s_{12})^{-3-2\e} \, S_\Gamma \, \frac{(1-x_1)^{-1-2\e}(1-x_2)^{-1-2\e}}
{2x_1} \Bigg[ \nonumber \\
&& \frac{3}{\e^2} + \frac{1}{\e} \left(
          - 4 \log 2
          + 4 H(-1;x_1)
          - 4 H(0;x_1)
         \right )
+ \bigg(
        2 \log^2 2
          - \frac{4\pi^2}{3} \nonumber \\ &&
          + \log 2 \left(
          - 4 H(-1;x_1) 
          + 4 H(0;x_1)\right) 
          + 4 H(-1,-1;x_1)
          - 4 H(-1,0;x_1)\nonumber \\ &&
          - 4 H(0,-1;x_1)
          + 4 H(0,0;x_1)
 \bigg)
       + \e  \bigg(
          - 14 \zeta_3
          + \frac{8\pi^2}{3} \left( \log 2
          -  H(-1;x_1)
          +  H(0;x_1) \right)\nonumber \\ &&
          + 4\log 2 \left(
          -  H(-1,-1;x_1) 
          -  H(-1,1;x_1) 
          +  H(0,-1;x_1)
          +  H(0,1;x_1) \right)\nonumber \\ &&
          + 4 H(-1,-1,-1;x_1)
          - 4 H(-1,-1,0;x_1)
          + 4 H(-1,1,-1;x_1)
          - 4 H(-1,1,0;x_1)\nonumber \\ &&
          - 4 H(0,-1,-1;x_1)
          + 4 H(0,-1,0;x_1)
          - 4 H(0,1,-1;x_1)
          + 4 H(0,1,0;x_1)
          \bigg)\nonumber \\ &&
          + {\cal O} \big(  (1-x_2) \big)
\Bigg] + {\cal O} (\e^2)\;,
\end{eqnarray}

\subsection{Collinear $x_2$ region}
In the collinear $x_2$ region, we can expand around $x_1=1$, and the master integrals read:
\begin{eqnarray}
 I_{13} & = & 
(s_{12})^{-3-2\e} \, S_\Gamma \, \frac{(1-x_1)^{-2\e}(1-x_2)^{-2\e}}
{1+2 x_2} \Bigg[ \nonumber \\
&&       \frac{1}{\e} \left(
          - \log 2
          + H(-1;x_2)
          - 2H(0;x_2)
          \right)\nonumber \\ &&
       + \bigg( \frac{3}{2}\log^2 2
          +\log 2\left(- H(-1;x_2)
          + 4H(0;x_2)
          + 2H(1;x_2)\right)
          + \frac{2\pi^2}{3}
          + H(-1,-1;x_2)\nonumber \\ &&
          - 2H(-1,0;x_2)
          - 4H(0,-1;x_2)
          + 6H(0,0;x_2)
          - 2H(1,-1;x_2)
          + 4H(1,0;x_2)\bigg)\nonumber \\ &&
       + \e  \bigg(
           4\zeta_3
          - \frac{7}{6}\log^3 2
          + \log^2 2 \left(\frac{1}{2} H(-1;x_2)
          - 4H(0;x_2)
          - 3H(1;x_2)\right)\nonumber \\ &&
	+\log 2 \big(
          - H(-1,-1;x_2)
          + 2H(-1,0;x_2)
          + 4H(0,-1;x_2)
          - 10H(0,0;x_2)\nonumber \\ &&
          - 4H(0,1;x_2)
          + 2H(1,-1;x_2)
          - 8H(1,0;x_2)
          - 4H(1,1;x_2)\big) \nonumber \\ &&
          - \frac{2\pi^2}{3} \left( \log 2
           + H(0;x_2)
           + 2 H(1;x_2)\right)
          + H(-1,-1,-1;x_2)
          - 2H(-1,-1,0;x_2)\nonumber \\ &&
          - 2H(-1,0,-1;x_2)
          + 4H(-1,0,0;x_2)
          + 2H(-1,1,0;x_2)
          - 4H(0,-1,-1;x_2)\nonumber \\ &&
          + 6H(0,-1,0;x_2)
          + 10H(0,0,-1;x_2)
          - 14H(0,0,0;x_2)
          + 4H(0,1,-1;x_2)\nonumber \\ &&
          - 8H(0,1,0;x_2)
          - 2H(1,-1,-1;x_2)
          + 4H(1,-1,0;x_2)
          + 8H(1,0,-1;x_2)\nonumber \\ &&
          - 12H(1,0,0;x_2)
          + 4H(1,1,-1;x_2)
          - 8H(1,1,0;x_2)
          \bigg) \nonumber \\ 
      &&    + {\cal O} \big(  (1-x_1) \big)
\Bigg] + {\cal O} (\e^2)\;,\\
 I_{16} & = & 
(s_{12})^{-3-2\e} \, S_\Gamma \, \frac{(1-x_1)^{-2\e}(1-x_2)^{-2\e}}
{1-x_2^2} \Bigg[ \nonumber \\
&&
       -\frac{1}{\e^2}
       + \frac{1}{\e} \left(
           \log 2
          - H(-1;x_2)
          \right)
+ \bigg(
       - \frac{1}{2}\log^2 2
          + H(-1;x_2)\log 2
          + \frac{2\pi^2}{3}\nonumber \\ &&
          - H(-1,-1;x_2)
          + 2H(0,0;x_2)
          + 2H(1,0;x_2)\bigg)
       + \e  (
           6\zeta_3
          + \frac{1}{6}\log^3 2\nonumber \\ &&
          - \frac{1}{2}\, \log^2 2 H(-1;x_2)
          + \log 2 \left(H(-1,-1;x_2)
          - 2H(0,0;x_2)
          - 2H(1,0;x_2)\right) \nonumber \\ &&
          + \frac{2\pi^2}{3} \left(
          - \log 2
          + H(-1;x_2)
          - H(0;x_2)
          - H(1;x_2)\right)
          - H(-1,-1,-1;x_2)\nonumber \\ &&
          + 2H(-1,0,0;x_2)
          + 2H(-1,1,0;x_2)
          + 2H(0,-1,0;x_2)
          + 2H(0,0,-1;x_2)\nonumber \\ &&
          - 6H(0,0,0;x_2)
          - 4H(0,1,0;x_2)
          + 2H(1,-1,0;x_2)
          + 2H(1,0,-1;x_2)\nonumber \\ &&
          - 6H(1,0,0;x_2)
          - 4H(1,1,0;x_2)
          )
   + {\cal O} \big(  (1-x_1) \big)
\Bigg] + {\cal O} (\e^2)\;,\\
 I_{17} & = & 
(s_{12})^{-3-2\e} \, S_\Gamma \, \frac{(1-x_1)^{-1-2\e}(1-x_2)^{-2\e}}
{2 x_2^2} \Bigg[ \nonumber \\
&&
       -\frac{1}{\e^2}
       + \frac{2}{\e}  H(0;x_2)
+ \bigg(
        2\log^2 2
          + 4H(1;x_2)\log 2
          - \frac{\pi^2}{3}
          - 4H(0,0;x_2)
          - 4H(1,-1;x_2)\nonumber \\ &&
          - 2H(1,0;x_2)
\bigg)
       + \e  \bigg(
          - 4\zeta_3
          - \frac{8}{3}\log^3 2
         + \log^2 2 \left(
          - 4H(0;x_2)
          - 8H(1;x_2)\right) \nonumber \\ &&
	+ \log 2 \left(
          - 8H(0,1;x_2)
          + 4H(1,-1;x_2)
          - 4H(1,0;x_2)
          - 12H(1,1;x_2)\right) \nonumber \\ &&
          + \frac{\pi^2}{3}\left(2\log 2
          + 2H(0;x_2)
          + 5H(1;x_2)\right)
          + 8H(0,0,0;x_2)
          + 8H(0,1,-1;x_2)\nonumber \\ &&
          + 4H(0,1,0;x_2)
          - 4H(1,-1,-1;x_2)
          + 4H(1,-1,0;x_2)
          + 4H(1,0,-1;x_2)\nonumber \\ &&
          + 6H(1,0,0;x_2)
          + 12H(1,1,-1;x_2)
          + 4H(1,1,0;x_2)
          \bigg)
\nonumber \\
      &&    + {\cal O} \big(  (1-x_1) \big)
\Bigg] + {\cal O} (\e^2)\;,\\
 M_{22} & = & 
(s_{12})^{-3-2\e} \, S_\Gamma \, \frac{(1-x_1)^{-2\e}(1-x_2)^{-2\e}}
{x_2} \Bigg[
       \frac{1}{2\e^2}
       + \frac{1}{2\e} \left(
          - \log 2
          +H(-1;x_2)
          \right)\nonumber \\ &&
          + \frac{1}{4} \log^2 2
          - \frac{1}{2} H(-1;x_2) \log 2
          - \frac{\pi^2}{3}
          + \frac{1}{2} H(-1,-1;x_2)
          - H(0,0;x_2)
          - H(1,0;x_2)\nonumber \\ &&
       + \e \Bigg(
          - 3 \zeta_3
          - \frac{1}{12} \log^3 2
          + \frac{1}{4}\log^2 2\, H(-1;x_2) \nonumber \\ &&
          + \log 2 \left( 
          - \frac{1}{2} H(-1,-1;x_2)
          + H(0,0;x_2)
          + H(1,0;x_2)\right)\nonumber \\ &&
          + \frac{\pi^2}{3} \left(\log 2
          -  H(-1;x_2)
          +  H(1;x_2)
          +  H(0;x_2)\right)
          + \frac{1}{2} H(-1,-1,-1;x_2)\nonumber \\ &&
          - H(-1,0,0;x_2)
          - H(-1,1,0;x_2)
          - H(0,-1,0;x_2)
          - H(0,0,-1;x_2)\nonumber \\ &&
          + 3 H(0,0,0;x_2)
          + 2 H(0,1,0;x_2)
          - H(1,-1,0;x_2)
          - H(1,0,-1;x_2)\nonumber \\ &&
          + 3 H(1,0,0;x_2)
          + 2 H(1,1,0;x_2)
          \Bigg)
   + {\cal O} \big(  (1-x_1) \big)
\Bigg] + {\cal O} (\e^2)\;,\\
 I_{23} & = & 
(s_{12})^{-3-2\e} \, S_\Gamma \, \frac{(1-x_1)^{-1-2\e}(1-x_2)^{-2\e}}
{2x_2(1+x_2)} \Bigg[ \nonumber \\
  &&
       \frac{2}{\e}  
          H(0;x_2)          
       - \frac{4\pi^2}{3}
          - 4H(0;x_2)\log 2
          - 4H(-1,0;x_2)
          + 4H(0,-1;x_2)
          - 6H(0,0;x_2)\nonumber \\ &&
          - 4H(1,0;x_2)
          + 4g_{23b}(x_2)    + {\cal O} \big(  (1-x_1) \big)
\Bigg] + {\cal O} (\e)\;,\\
 I_{24} & = & 
(s_{12})^{-3-2\e} \, S_\Gamma \, \frac{(1-x_1)^{-2\e}(1-x_2)^{-1-2\e}}
{x_2} \Bigg[ \nonumber \\
 &&
         \frac{1}{\e}  (
          - \log 2
          + H(-1;x_2)
          )
+ \bigg(
        \frac{3}{2}\log^2 2
          - \frac{\pi^2}{6}
          + \log 2  (-H(-1;x_2)
          + 2H(1;x_2))\nonumber \\ &&
           + H(-1,-1;x_2)
          - 2H(1,-1;x_2)\bigg)
       + \e  \bigg(
          - \frac{13}{4}\zeta_3
          - \frac{7}{6}\log^3 2 \nonumber \\ &&
          + \frac{1}{2} \log^2 2 (H(-1;x_2)
          - 6H(1;x_2))
	+ \log 2\big(
           2H(1,0;x_2)
          + 2H(0,0;x_2)
                    - 4H(1,1;x_2)\nonumber \\ &&
          + 2H(1,-1;x_2)
          - H(-1,-1;x_2)
         \big)
          + \frac{\pi^2}{3}\left( 3 \log 2
          - 2H(-1;x_2)
          + H(1;x_2)\right)\nonumber \\ &&
          + H(-1,-1,-1;x_2)
          - 2H(-1,0,0;x_2)
          - 2H(-1,1,0;x_2)
          - 2H(0,-1,0;x_2)\nonumber \\ &&
          - 2H(0,0,-1;x_2)
          - 2H(1,-1,-1;x_2)
          - 2H(1,-1,0;x_2)
          - 2H(1,0,-1;x_2)\nonumber \\ &&
          + 4H(1,1,-1;x_2)
          \bigg)
        + {\cal O} \big(  (1-x_1) \big)
\Bigg] + {\cal O} (\e^2)\;,\\
 I_{25} & = & 
(s_{12})^{-3-2\e} \, S_\Gamma \, \frac{(1-x_1)^{-1-2\e}(1-x_2)^{-2\e}}
{2x_2} \Bigg[ \nonumber \\
&& 
       \frac{2}{\e}   H(0;x_2)
       - \pi^2
          - 4H(0;x_2)\log 2
          + 4H(0,-1;x_2)
          - 8H(0,0;x_2)\nonumber \\ &&
          - 4H(1,0;x_2)
          + 4g_{23b}(x_2)
    + {\cal O} \big(  (1-x_1) \big)
\Bigg] + {\cal O} (\e)\;,\\
I_{27} & = & 
(s_{12})^{-3-2\e} \, S_\Gamma \, \frac{(1-x_1)^{-1-2\e}(1-x_2)^{-2\e}}
{2x_2} \Bigg[ \nonumber \\
  &&
         -\frac{1}{\e^2}
          +\frac{1}{\e}  (
           2\log 2
          - 2H(-1;x_2)
          )
+\big(
       - 2\log^2 2
          - 4\log 2 H(1;x_2)
          + \pi^2
          + 2H(0,0;x_2)\nonumber \\ &&
          + 4H(1,-1;x_2)
          + 2H(1,0;x_2)
\big)
       + \e  \bigg(
           \frac{25}{2}\zeta_3
          + \frac{4}{3}\log^3 2
          + \log^2 2 (2H(-1;x_2)\nonumber \\ &&
          + 6H(1;x_2))
          + \log 2 \big( 4H(-1,0;x_2)
          + 4H(-1,1;x_2)
                    - 4H(0,0;x_2)
          + 12H(1,1;x_2)\big)\nonumber \\ && 
          + \frac{\pi^2}{3} \left(
          - 6\log 2
          + 4 H(-1;x_2)
          - 3 H(0;x_2)
          - 5H(1;x_2)\right)
          - 4H(-1,0,-1;x_2)\nonumber \\ &&
          + 4H(-1,0,0;x_2)
          - 4H(-1,1,-1;x_2)
          + 4H(-1,1,0;x_2)
          + 4H(0,0,-1;x_2)\nonumber \\ &&
          - 6H(0,0,0;x_2)
          - 4H(0,1,0;x_2)
          - 6H(1,0,0;x_2)
          - 12H(1,1,-1;x_2)\nonumber \\ &&
          - 4H(1,1,0;x_2)
          \bigg)
    + {\cal O} \big(  (1-x_1) \big)
\Bigg] + {\cal O} (\e^2)\;,\\
I_{28} & = & 
(s_{12})^{-3-2\e} \, S_\Gamma \, \frac{(1-x_1)^{-1-2\e}(1-x_2)^{-2\e}}
{2x_2} \Bigg[ \nonumber \\
 && 
         -\frac{2}{\e^2}
       + \frac{1}{\e}  (
           2\log 2
          - 2H(-1;x_2)
          + 2H(0;x_2)
          )
       + \frac{2\pi^2}{3}
       + \e  \bigg(
           4\zeta_3
          - \frac{4}{3}\log^3 2\nonumber \\ &&
          + \log^2 2 \left( 2H(-1;x_2)
          - 4H(0;x_2)
          - 2H(1;x_2)\right)
          + \log 2 \big( 4H(-1,0;x_2)\nonumber \\ &&
          + 4H(-1,1;x_2)
          + 4H(0,-1;x_2)
          - 8H(0,0;x_2)
          - 4H(0,1;x_2)
          + 4H(1,-1;x_2)\nonumber \\ &&
          - 4H(1,0;x_2)\big) 
	+ \frac{4\pi^2}{3} \left (
          - \log 2
          + H(-1;x_2)
          - H(0;x_2)\right)
          - 4H(-1,0,-1;x_2)\nonumber \\ &&
          + 4H(-1,0,0;x_2)
          - 4H(-1,1,-1;x_2)
          + 4H(-1,1,0;x_2)
          - 4H(0,-1,-1;x_2)\nonumber \\ &&
          + 4H(0,-1,0;x_2)
          + 8H(0,0,-1;x_2)
          - 8H(0,0,0;x_2)
          + 4H(0,1,-1;x_2)\nonumber \\ &&
          - 4H(0,1,0;x_2)
          - 4H(1,-1,-1;x_2)
          + 4H(1,-1,0;x_2)
          + 4H(1,0,-1;x_2)\nonumber \\ &&
          - 4H(1,0,0;x_2)
          \bigg)
      + {\cal O} \big(  (1-x_1) \big)
\Bigg] + {\cal O} (\e^2)\;,\\
I_{29} & = & 
(s_{12})^{-3-2\e} \, S_\Gamma \, \frac{(1-x_1)^{-1-2\e}(1-x_2)^{-1-2\e}}
{2x_2} \Bigg[ \nonumber \\
&& \frac{3}{\e^2}  + \frac{1}{\e} \left(
          - 4 \log 2
          + 4 H(-1;x_2 )
          - 4 H(0;x_2 )
          \right)
       + \bigg( 2 \log^2 2
          - \frac{4\pi^2}{3}\nonumber \\ &&
          + \log 2 \left(
          - 4 H(-1;x_2 )
          + 4 H(0;x_2 )\right)
          + 4 H(-1,-1;x_2 )
          - 4 H(-1,0;x_2 )\nonumber \\ &&
          - 4 H(0,-1;x_2 )
          + 4 H(0,0;x_2 )\bigg)
       + \e  \bigg(
          - 14 \zeta_3
          + \frac{8\pi^2}{3} \left( \log 2
          - H(-1;x_2 ) + H(0;x_2 ) \right)\nonumber \\ &&
          + \log 2\left(
          - 4 H(-1,-1;x_2 ) 
          - 4 H(-1,1;x_2 ) 
          + 4 H(0,-1;x_2 )
          + 4 H(0,1;x_2 )\right) \nonumber \\ &&
          + 4 H(-1,-1,-1;x_2 )
          - 4 H(-1,-1,0;x_2 )
          + 4 H(-1,1,-1;x_2 )
          - 4 H(-1,1,0;x_2 )\nonumber \\ &&
          - 4 H(0,-1,-1;x_2 )
          + 4 H(0,-1,0;x_2 )
          - 4 H(0,1,-1;x_2 )
          + 4 H(0,1,0;x_2 )
          \bigg)\nonumber \\ &&
  + {\cal O} \big(  (1-x_1) \big) 
\Bigg] + {\cal O} (\e^2)\;.
\end{eqnarray}
The combination $(1+x_2)I_{23}-I_{25}$ appears in this collinear limit to 
order ${\cal O}(\e)$, it reads:
\begin{eqnarray}
(1+x_2)I_{23}-I_{25} & = & 
(s_{12})^{-3-2\e} \, S_\Gamma \, \frac{(1-x_1)^{-1-2\e}(1-x_2)^{-2\e}}
{2x_2} \Bigg[ \nonumber \\
&& 
       - \frac{\pi^2}{3}
          - 4 H(-1,0;x_2)
          + 2 H(0,0;x_2)\nonumber \\ &&
       + \e \Bigg(
          - 30 \zeta_3
          + \log 2 \left(
           8 H(-1,0;x_2)
          - 4 H(0,0;x_2)\right)\nonumber \\ &&
         + \frac{2\pi^2}{3}\left(
           \log 2
          + 4 H(-1;x_2)
          - 2 H(0;x_2)
          +  H(1;x_2)\right)\nonumber \\ &&
          + 8 H(-1,-1,0;x_2)
          - 8 H(-1,0,-1;x_2)
          + 12 H(-1,0,0;x_2)\nonumber \\ &&
          + 8 H(-1,1,0;x_2)
          - 4 H(0,-1,0;x_2)
          + 4 H(0,0,-1;x_2)\nonumber \\ &&
          - 10 H(0,0,0;x_2)
          - 4 H(0,1,0;x_2)
          + 8 H(1,-1,0;x_2)\nonumber \\ &&
          - 4 H(1,0,0;x_2)
          + 8 g_{23c}(x_2)
          \Bigg)
    + {\cal O} \big(  (1-x_1) \big)
\Bigg] + {\cal O} (\e^2)\;,
\end{eqnarray}

\subsection{Soft region}
In the soft region, $x_1\to 1$ and $x_2\to 1$. The master integrals then become:
\begin{eqnarray}
I_{13} & = & 
(s_{12})^{-2-2\e} \, S_\Gamma \, \frac{(1-x_1)^{-2\e}(1-x_2)^{1-2\e}}
{2\e(1-2\e)} \left[ 1 -\frac{\pi^2}{3} \e^2 - 4 \zeta_3 \e^3 - \frac{\pi^4}{45} \e^4 + {\cal O} (\e^5)   \right] \;, \nonumber \\
I_{16} & = & 
-(s_{12})^{-3-2\e} \, S_\Gamma \, \frac{(1-x_1)^{-2\e}(-1-x_2)^{1-2\e}}
{2\e^2} \left[ 1 -\frac{\pi^2}{3} \e^2 - 4 \zeta_3 \e^3 - \frac{\pi^4}{45} \e^4 + {\cal O} (\e^5)   \right] \;, \nonumber \\
I_{17} & = & 
-(s_{12})^{-3-2\e} \, S_\Gamma \, \frac{(1-x_1)^{-1-2\e}(1-x_2)^{-2\e}}
{2\e^2} \left[ 1 -\frac{\pi^2}{3} \e^2 - 4 \zeta_3 \e^3 - \frac{\pi^4}{45} \e^4 + {\cal O} (\e^5)   \right] \;, \nonumber \\
I_{22} & = & 
(s_{12})^{-3-2\e} \, S_\Gamma \, \frac{(1-x_1)^{-2\e}(1-x_2)^{-2\e}}
{2\e^2} \left[ 1 -\frac{\pi^2}{3} \e^2 - 4 \zeta_3 \e^3 - \frac{\pi^4}{45} \e^4 + {\cal O} (\e^5)   \right] \;, \nonumber \\
I_{23} & = & 
-(s_{12})^{-3-2\e} \, S_\Gamma \, \frac{(1-x_1)^{-1-2\e}(1-x_2)^{1-2\e}}
{2\e(1-2\e)} \left[ 1 -\frac{\pi^2}{3} \e^2 - 4 \zeta_3 \e^3 - \frac{\pi^4}{45} \e^4 + {\cal O} (\e^5)   \right] \;, \nonumber \\
I_{24} & = & 
(s_{12})^{-3-2\e} \, S_\Gamma \, \frac{(1-x_1)^{-2\e}(1-x_2)^{-2\e}}
{2\e} \left[ 1 -\frac{\pi^2}{3} \e^2 - 4 \zeta_3 \e^3 - \frac{\pi^4}{45} \e^4 + {\cal O} (\e^5)   \right] \;, \nonumber \\
I_{25} & = & 
- (s_{12})^{-3-2\e} \, S_\Gamma \, \frac{(1-x_1)^{-1-2\e}(1-x_2)^{1-2\e}}
{\e(1-2\e)} \left[ 1 -\frac{\pi^2}{3} \e^2 - 4 \zeta_3 \e^3 - \frac{\pi^4}{45} \e^4 + {\cal O} (\e^5)   \right] \;, \nonumber \\
I_{27} & = & 
-(s_{12})^{-3-2\e} \, S_\Gamma \, \frac{(1-x_1)^{-1-2\e}(1-x_2)^{-2\e}}
{2\e^2} \left[ 1 -\frac{\pi^2}{3} \e^2 - 4 \zeta_3 \e^3 - \frac{\pi^4}{45} \e^4 + {\cal O} (\e^5)   \right] \;, \nonumber \\
I_{28} & = & 
-(s_{12})^{-3-2\e} \, S_\Gamma \, \frac{(1-x_1)^{-1-2\e}(1-x_2)^{-2\e}}
{\e^2} \left[ 1 -\frac{\pi^2}{3} \e^2 - 4 \zeta_3 \e^3 - \frac{\pi^4}{45} \e^4 + {\cal O} (\e^5)   \right] \;, \nonumber \\
I_{29} & = & 
(s_{12})^{-3-2\e} \, S_\Gamma \,  \frac{(1-x_1)^{-1-2\e}(1-x_2)^{-1-2\e}}{\e^2}
 \left[ \frac{3}{2} -\frac{2\pi^2}{3}\e^2 - 11\zeta_3 \e^3 - \frac{3\pi^4}{20} \e^4  + {\cal O} (\e^5)   \right] \;.\nonumber \\ &&
\end{eqnarray}

\end{document}